\documentclass[journal=jacsat,manuscript=article]{achemso}
\pdfoutput=1
\usepackage[version=3]{mhchem} 
\usepackage{epstopdf}
\usepackage{color}
\usepackage[exponent-product=\cdot,per-mode=symbol]{siunitx}

\author{Christian Appel}
\email{christian.appel@fkp.physik.tu-darmstadt.de}
\author{Bj\"orn Kuttich}
\affiliation[Institute of Condensed Matter Physics]
{Institute of Condensed Matter Physics, Technische Universit\"at Darmstadt, \\ Hochschulstr. 8, D-64289 Darmstadt, Germany}
\author{Lukas St\"uhn}
\author{Robert W. Stark}
\affiliation[Physics of Surfaces]
{Physics of Surfaces, Technische Universit\"at Darmstadt, Alarich-Weiss-Str. 16, \\ D-64287 Darmstadt, Germany}
\author{Bernd St\"uhn}
\affiliation[Institute of Condensed Matter Physics]
{Institute of Condensed Matter Physics, Technische Universit\"at Darmstadt, \\ Hochschulstr. 8, D-64289 Darmstadt, Germany}

\title[An \textsf{achemso} demo]
{Zero Field Assembly of Long Magnetic Dipolar Chains in 2D Polymer Nanocomposite Films}
\abbreviations{}

\begin{document}
\section{Abstract}
The existence of magnetic dipolar nanoparticle chains at zero field has been predicted theoretically for decades, but these structures are rarely observed experimentally.
A prerequisite is a permanent magnetic moment on the particles forming the chain.
Here we report on the observation of magnetic dipolar chains of spherical iron oxide nanoparticles with a diameter of \SI{12.8}{\nano\meter}. The nanoparticles are embedded in an ultrathin polymer film.
Due to the high viscosity of the polymer matrix, the dominating aggregation mechanism is driven by dipolar interactions.
Smaller iron oxide nanoparticles (\SI{9.4}{\nano\meter})	show no permanent magnetic moment and do not form chains but compact aggregates.
Mixed monolayers of different iron oxide nanoparticles and polymer  at the air-water interface are characterized by Langmuir isotherms and in-situ X-ray reflectometry (XRR). 
The combination of the particles with a polymer leads to a stable polymer nanocomposite film at the air-water interface.
XRR experiments show that nanoparticles are immersed in a thin polymer matrix of \SI{3}{\nano\meter}.
Using atomic force microscopy (AFM) on Langmuir-Blodgett films, we measure the lateral distribution of particles in the film.
An analysis of single structures within transferred films results in fractal dimensions that are in excellent agreement with 2D simulations. 

\section{Introduction}
Due to their perfect combination of well adjustable macroscopic properties, polymer nanocomposites have attracted a lot of research interest in recent years. The polymeric component of the composites serves  as a matrix in which a wide range of nanoparticles can be introduced in order to achieve desired thermal, rheological, optical, conductive, magnetic, etc. properties.\cite{Kelly2003,Balazs2006,Winey2007,Hanemann2010,Kumar2017} While tuning of these properties is of outstanding interest in most applied research, fundamental studies focused on the precise interactions between polymer and particles as well as on the microstructure of the nanoparticles distributed in the polymer matrix.\cite{Chevigny2011,Chandran2014,Jiang2015,Starr2016}
To prevent nanoparticles from aggregation they are stabilised by a shell.\cite{Zhou2009}
The use of carboxyl acids, such as oleic acid, as shell ligands leads to a steric stabilisation of the particles. 
In a composite the shell type strongly influences the interaction between nanoparticles and polymer matrix.

Metal nanoparticles are not only characterized by their high surface to volume ratio but also by quantum confinement.\cite{White2009} Most prominent among these are gold nanoparticles, especially in combination with polymers, because of their localised surface plasmon resonance, allowing for many sensor applications.\cite{Kelly2003,Jain2008,Mayer2011} Besides these, nanoparticles composed of iron, nickel and cobalt have come more and more into focus because of their outstanding magnetic properties.\cite{LesliePelecky1996,Skomski2003,Gao2009} Sufficiently small particles (size $\approx\SI{100}{\nano\metre}$) consist only of one single ferromagnetic domain, while for even smaller particles the permanent magnetisation completely vanishes. The material is below its Curie temperature but its magnetic dipole moment flips on a fast timescale. This state is known as superparamagnetic.\cite{LesliePelecky1996,Bedanta2008,Koplovitz2019} Introducing magnetic particles into a non-magnetic solvent yields a so-called ferrofluid, which offers the possibility to investigate fundamental magnetic dipole-dipole interactions.\cite{Luo1991,Teixeira2000,Butter2003,Klokkenburg2006} A polymer nanocomposite consisting of magnetic nanoparticles can be understood as a ferrofluid-like system with a highly viscous solvent. Due to the high viscosity, Brownian dynamics are significantly slowed down. It may therefore be expected that aggregation of nanoparticles is unlikely to be limited by diffusion (DLA) but dominated by magnetic interactions.
This effect is particularly interesting when the system's degrees of freedom are reduced to two dimensions (2D). 
Firstly, the probability of direct particle interaction is increased due to the 2D confinement.
Secondly, aggregated structures can be visualized directly by real space methods.
At the air-water interface, polymer and nanoparticles can be prepared into an ultra thin film and investigated by various experimental methods.

In the following we will investigate the influence of particle size on the dipolar interactions and the impact on  structure formation of nanoparticles inside a polymer monolayer. We choose to study spherical iron oxide nanoparticles having sizes between \SI{5}{\nano\metre} and \SI{20}{\nano\metre} which are commercially available and should be close to the limiting size for expected superparamagnetic behaviour.\cite{Salazar2011,Carvalho2013,Gabbasov2015} They are stabilised by a hydrophobic oleic acid shell. The used polymer is a hydrophilic-hydrophobic diblock copolymer. For the preparation of the nanocomposite film we make use of the air-water interface and the fact that both nanoparticles and the used polymer are surface active and thus form stable monolayers at this interface.

The properties of the polymer at the air-water interface have been investigated recently.\cite{Appel2018} It forms a thin layer with thickness between $\SI{2}{\nano\meter}$ and  $\SI{3}{\nano\meter}$. In the present study we first characterise three iron oxide nanoparticles of different sizes by small angle X-ray scattering. We investigate their capability to form stable monolayers at the air-water interface using Langmuir isotherms and in-situ X-ray reflectivity (XRR).
Our main interest lies in the properties of the combined films of nanoparticles and polymer. 
First, the aforementioned experimental techniques are used to determine the properties of the composite films. We then turn to the question of the structure of nanoparticle aggregates forming within the polymer layer. 
Depending on the aggregation mechanism being diffusion limited (DLA) or dominated by magnetic dipole interaction, the appearance of differently formed clusters is predicted.\cite{Pastor1995,Vicsek1984}
We use atomic force microscopy (AFM) on ex-situ samples (Langmuir-Blodgett (LB) films) to answer this question.

\section{Experimental Section}
\label{sec:experimental_section}
\subsection{Samples}
Samples used in this work are three different sizes of iron oxide nanoparticles (diameter: 5, 10 and \SI{20}{\nano\metre}) and a diblock copolymer poly(ethylene glycol)-\textit{b}-poly(\textit{n}-butyl acrylate) (PEG$_6$-\textit{b}-P\textit{n}BA$_{132}$, M$_\text{n}=\SI{17.2}{\kilo\gram\per\mol}$). The polymer was synthesized via controlled living polymerization as described in our previous publication.\cite{Appel2018}
The iron oxide nanoparticles were purchased from \textit{Ocean Nanotech} where their size and shape was characterized by TEM experiments.
The \SI{5}{\nano\metre} particles (FeNP5) are stated to consist of a mixture of Fe$_2$O$_3$ and Fe$_3$O$_4$, while the \SI{10}{\nano\metre} (FeNP10) and \SI{20}{\nano\metre} (FeNP20) ones are claimed to only consist of Fe$_3$O$_4$ nanocrystals.
Particles were purchased dissolved in chloroform and stabilized by an oleic acid shell of \SI{2.5}{\nano\metre} thickness according to the supplier.

\subsection{Pressure-Area Isotherms}
Pressure-area isotherms were performed on a KSV NIMA trough system (KSV NIMA, Langmuir Troughs KN 1006, Biolin Scientific) equipped with two hydrophobic symmetrically-moving barriers and a Wilhelmy platinum plate placed exactly in-between those barriers (\SI{45}{\degree}- tilted with respect to the direction of compression) at room temperature of $22\pm\SI{1}{\celsius}$.
The trough dimensions are $\left(58\hspace{0.8mm}x\hspace{0.8mm}14.5\right)\,\si{\centi\metre\squared}$.
For each measurement both barriers and the trough are first cleaned with ethanol and rinsed with purified water.
We use deionized Millipore water (Millipore Direct-Q) with a specific resistivity of \SI{18.2}{\mega\ohm\centi\metre} at \SI{25}{\celsius} for rinsing and for the subphase.
All isotherms were measured on a water surface and the purity of the surface was checked prior to each measurement by a compression of the water surface with \SI{2175}{\milli\metre\squared\per\minute} while the surface pressure was monitored ($\Delta \Pi < \SI{0.3}{\milli\newton\per\metre}$). 
The iron oxide nanoparticles and composite samples were dissolved in chloroform (concentration $c \approx \SI{5.0}{\milli\gram\per\milli\litre}$) and mixed before each measurement (Vortex 2000 mixer).
For the composite samples the concentrations were $c_{\text{PnBA}} = \SI{1}{\milli\gram\per\milli\litre} : c_{\text{FeNP}} =1...\SI{3}{\milli\gram\per\milli\litre}$ so that area fractions of nanoparticles spread at the interface are comparable for both systems ($\SI{3}{\percent}$ particle coverage at maximum trough area).
A specified volume $V$ was spread in drops of \SI{1.5}{\micro\litre} on the water subphase with a Hamilton syringe (maximum volume \SI{5}{\micro\litre}).  
The chloroform evaporated quickly and the composite formed stable monolayers.
All isotherms were measured after a waiting time of \SI{30}{min}.
The available surface for each object (i.e. nanoparticle or polymer molecule) is given by the mean molecular area mmA, which is defined as:
\begin{equation}
	\text{mmA} = \frac{AM}{cV \mathrm{N_A}}
\end{equation}
$A$ is the actual area enclosed by the two barriers, $M$ the molecular weight of the object and $\mathrm{N_A}$ the Avogadro constant.
In case of the polymer nanocomposites, mmA was calculated for the polymer chain solely in order to compare to the pure polymer isotherm.
We subtracted the area occupied by nanoparticles from the area enclosed by the two barriers ($A$) to get the correct mmA for a polymer chain. 

The second parameter is the surface pressure $\Pi$ given by
\begin{equation}
	\Pi = \gamma_0 - \gamma
\end{equation}
where $\gamma_0$ is the surface tension of the subphase (water) and $\gamma$ the surface tension of the system after spreading the sample.
The layers were compressed with a constant velocity of $435\,\mathrm{mm }^2\,\mathrm{min}^{-1}$.
Isotherms were repeated at least three times to check for the  reproducibility of the results.

\subsection{X-ray Reflectometry}
Experiments were performed on a modified D8 Advance reflectometer (Bruker AXS, Germany).
The design of the setup enables us to measure reflectivities in $\theta-\theta$ geometry with the X-ray source and detector attached at goniometer arms which can be moved independently with a precision of $0.001^{\circ}$.
A conventional X-ray tube with a Cu anode (CuK$_{\alpha}$, wavelength $\lambda = 1.54 \,\text{\AA}$) is used to generate a X-ray beam with a line focus.
The beam is further monochromized by  a Goebel mirror (W/Si multilayer mirror).
Through a narrow horizontal slit of $0.1\,\mathrm{mm}$, the beam passes an absorber (calibrated Cu attenuator) which is necessary for high intensities near the critical angle in order to maintain a linear response of the detector.
A second $0.1\,\mathrm{mm}$ horizontal slit is placed after the absorber to cut out the K$_{\beta}$-line (which is also reflected by the monochromator).
Soller slits ($\Delta \theta_{\text{x}} = 25\,\mathrm{mrad}$) are placed after the last horizontal slit and directly in front of the detector.

Intensity is detected by a V\r{a}ntec-1 line detector (Bruker AXS, Germany) providing the possibility to measure the specularly reflected intensity and the diffuse intensity simultaneously in an angular range of $\Delta \theta_f = 2^{\circ}$ for a given incident angle $\theta_i$.
For each incident angle $\theta_i$, a single intensity I($\theta_f$) contains the specular and off-specular scattering.
The intensity of the reflected beam is determined as the integral over the specular peak corrected for background measured as the diffuse intensity.
The specular reflectivities were analyzed using the Motofit Reflectometry package, rev. 409\cite{Nelson2006} for IGOR Pro.

\subsection{Langmuir-Blodgett Films}
LB films were produced on a Langmuir trough system ($\mu$Trough System Kibron Inc., Helsinki Finland).
The silicon wafers were cleaned for 2 minutes in piranha solution and their surface was characterized by XRR measurements prior to the transfer of the monolayers.
For the transfer, the silicon wafers were first dipped through the monolayer in the subphase with a dipping speed of $2\,\mathrm{mm/min}$, and then pulled out of the subphase through the monolayer again at the same speed.
The molecular area of the film was controlled and monitored by a software to guarantee a constant pressure during the transfer.
The data shows that the monolayer was transferred when the silicon wafer was pulled out of the subphase (Z-type LB film) and all transfer ratios were $80-100\,$\%.

\subsection{Small Angle X-ray Scattering}
SAXS was performed using a laboratory X-ray set-up. 
The $K_{\alpha}$-line of a conventional copper X-ray tube with a wave length of $\lambda = 1.54\,\text{\AA}$ is used and further monochromated by a X-ray mirror.
The point focused beam is collimated by three pinholes.
The detector-sample distance is $1.5\,\mathrm{m}$ and calibration of the scattering vector $q$ is done by measuring the first peak of silver behenate as a calibration sample.\cite{Huang1993}
The scattering vector is defined as $|\vec{q}| = 4\pi/\lambda \, \sin\theta$, with $2\theta$ being the scattering angle. 
The scattered intensities are measured with a two-dimensional multi-wire gas detector (Molecular Metrology, $1024\,x\,1024$ pixels).
The overall instrumental resolution is estimated to be $\Delta \text{q} = 0.01\,\text{\AA}^{-1}$ by the full width half maximum of the silver behenate peak during calibration.
The accessible $q$ range is $0.007\,\text{\AA}^{-1} \leq q \leq 0.25\,\text{\AA}^{-1}$ and the data was radially averaged because all scattering was isotropic.

\subsection{Atomic Force Microscopy}
AFM images of polymer nanocomposite LB films were measured with a Cypher S AFM (Asylum Research Oxford Instruments) in amplitude modulation mode. 
Cantilevers (BudgetSensors and Olympus) with spring constants between $\SI{2}{\newton\per\meter}$ and $\SI{3}{\newton\per\meter}$, resonance frequencies between $\SI{70}{\kilo\hertz}$ and $\SI{75}{\kilo\hertz}$ and tip radii smaller than $\SI{10}{\nano\meter}$ were used. 
As feedback signal, the ratio of setpoint and free amplitude was kept at $0.6$ with a free amplitude of about $\SI{30}{\nano\meter}$.
Scan rates varied between $0.2$ and $2$ lines per second, depending on the scan size.
Measurements were performed in air under ambient conditions.
To correct for scanner drift, height images were second order flattened using the free software Gwyddion.\cite{Necas2012} 
For further analysis, nanoparticles were identified via a height threshold of $\SI{4}{\nano\meter}$.

\section{Properties of Iron Oxide Nanoparticles}

\subsection{Size and Shape of the Nanoparticles}

\begin{figure}[!t]
\centering
	\begin{small}
	\begin{tabular}{l c c c c c}
	\hline
	sample name & fit type & diameter$_\text{major}$ & diameter$_\text{minor}$ & PDI  \\ \hline
	FeNP$_{5}$  &spheroid & $9.4\pm 0.1\,\mathrm{nm}$ & $3.7 \pm 0.1\,\mathrm{nm}$& - \\ 
	FeNP$_{10}$ &sphere &$12.8\pm 0.1\,\mathrm{nm}$& $12.8\pm 0.1\,\mathrm{nm}$ & $0.10 \pm 0.03$ \\ 
	FeNP$_{20}$ &sphere & $21.4\pm 0.1\,\mathrm{nm}$  & $21.4\pm 0.1\,\mathrm{nm}$ & $0.11 \pm 0.02$ \\ \hline
	\end{tabular}
	\end{small}

\vspace{5mm}
	\centering
		\includegraphics[width=0.6\textwidth]{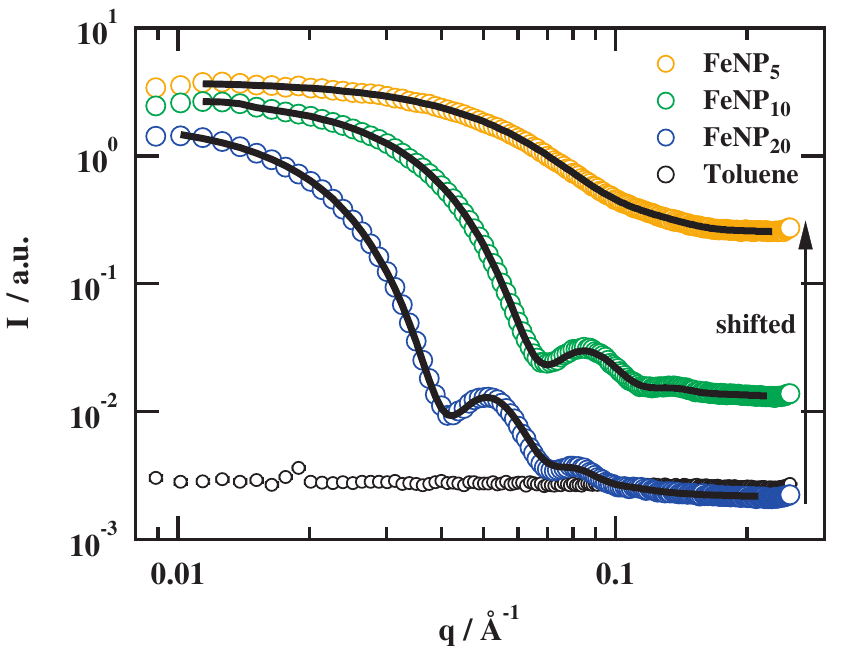}
	\caption{Small angle scattering data from the three different iron oxide nanoparticle sizes (FeNP5, FeNP10 and FeNP20). Very good agreement between fits and raw data is demonstrated using a polydisperse form factor of a sphere (FeNP10 and FeNP20) and a spheroid (FeNP5). The fit parameters for the models are summarized in the table above the figure.}
	\label{fig:fenp_saxs_formfactor} 
\end{figure}

Small angle X-ray scattering was performed on nanoparticle bulk solutions.
The solvent was changed to toluene in order to avoid the strong X-ray absorbance of chloroform and diluted to low concentrations ($c = 1\,\mathrm{mg/ml}$).
Resulting scattering profiles are shown in figure \ref{fig:fenp_saxs_formfactor}.
In all cases, the scattering at small q displays a plateau indicating single particle scattering. 
The scattering from FeNP5 is fundamentally different compared to the other two sizes.
For FeNP10 and FeNP20, clear oscillations are visible and intensity decays much faster ($\text{I} \sim \text{q}^{-4}$) compared to the small ones ($\text{I} \sim \text{q}^{-2}$).
The scattering from the large particles can be described by a spherical form factor with a polydisperse radius (polydispersity: PDI, details see ESI).
This is not possible for FeNP5, due to the weaker decaying intensity. Scattering from the smallest particles is modelled by an oblate spheroid form factor averaged over all possible orientations.
Thus, the described objects are anisotropic with a major and minor diameter. 
For these particles it is possible to determine a radius polydispersity due to the orientation average.
The oleic acid shell is not considered in the fitting, due to the missing contrast between oleic acid and toluene. 
Details on both form factors are given in the electronic supplementary information (ESI). 

For all particle sizes a very good agreement between model and experimental data is found as demonstrated by the full black lines in the figure.
The important parameters for data fitting are summarized in the table above figure \ref{fig:fenp_saxs_formfactor}. 
The results agree with the claimed sizes provided by \textit{Ocean Nanotech}.
However, the shape of FeNP5 particles is not spherical which leads to an additional degree of freedom for their orientation at the air-water interface. Since all small angle scattering data could be described by form factors only, the nanoparticles have to be dispersed homogeneously in the solvent. Thus, changing the solvent from chloroform to toluene did not lead to aggregation of the particles.

\subsection{Iron Oxide Nanoparticles at the Air-Water Interface}
Langmuir monolayers were prepared for all three particles sizes at low surface coverage.
Compression isotherms and \textit{in-situ} specular reflectivities were measured for the particle films.
Furthermore, the elastic modulus
\begin{equation}\label{eq:compressibility}
	\text{E} = -\text{mmA} \left(\frac{\partial \Pi}{\partial \text{mmA}}\right)
\end{equation}
was calculated from the isotherms as a measure for the resistance of the film upon compression.
No stable film is formed for FeNP20 particles.
Compression isotherms and specular reflectivities shown in the supporting information strongly indicate that FeNP20 particles aggregate after being spread on the air-water interface. Surface pressure is already significantly increased for mean molecular areas thirty times larger than the projected particle area measured by SAXS. XRR shows only a rough surface.
A similar behaviour is observed in a previous study.\cite{Vorobiev2015}
However, FeNP5 and FeNP10 form stable single layer particle films at the air-water interface.
In the following we will focus on the properties of these nanoparticles and their polymer composites.

\subsection{Langmuir Isotherms of Nanoparticle Layers} 

\begin{figure}[t]
	\centering
	\includegraphics[width=0.49\textwidth]{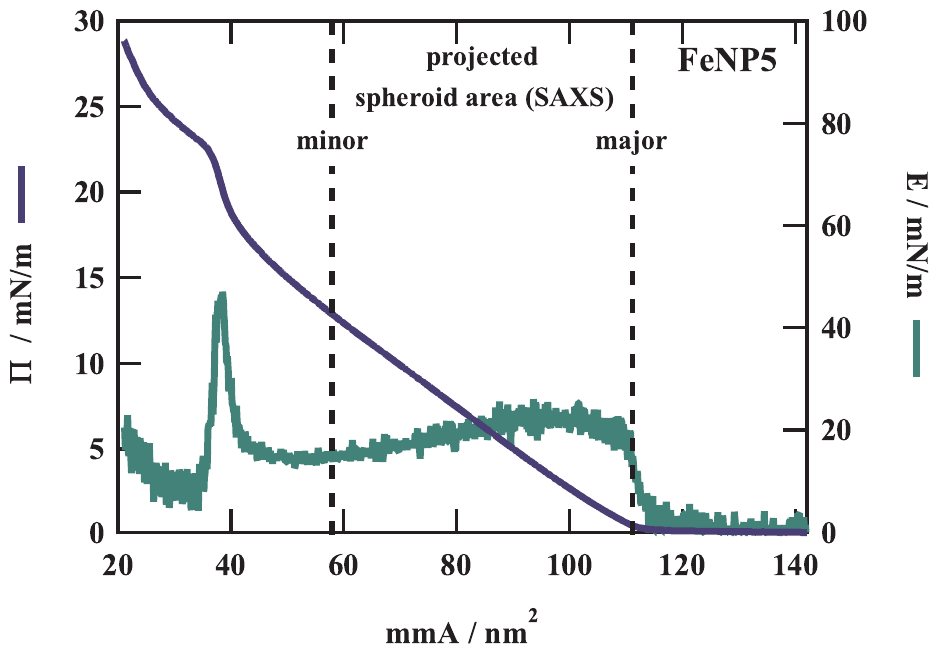}
	\hfill
	\includegraphics[width=0.49\textwidth]{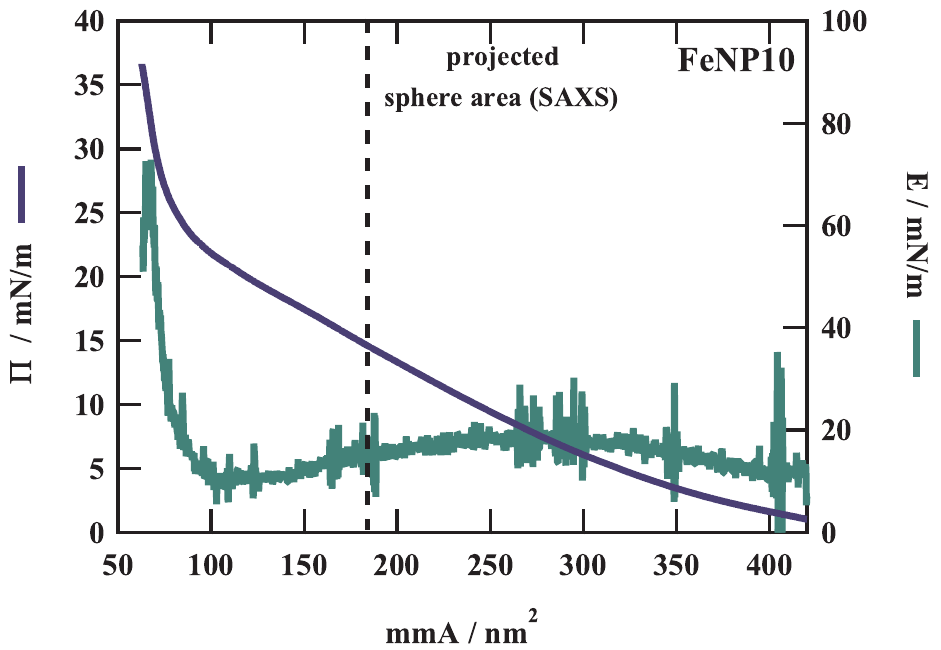}
	\caption{Langmuir isotherms presented as surface pressure ($\Pi$) vs. mean molecular area (mmA) for FeNP5 and FeNP10. The elastic modulus ($E$) computed from the isotherms is depicted as well (right axis). The vertical dashed lines represent the projected area for fully interpenetrating shells derived from the SAXS sizes (details see text).}
	\label{fig:isotherme_fenp5}
\end{figure}

We start with the compression behaviour of single particle films of FeNP5 and FeNP10 prepared at the air-water interface. 
The isotherm and elastic modulus of FeNP5 films is shown in the left panel of figure \ref{fig:isotherme_fenp5}.
For low surface coverage, the surface pressure remained low ($\Pi < \SI{1}{\milli\newton\per\metre}$) not affected by compression of the film.
The film can be easily compressed ($\text{E} \approx \SI{2}{\milli\newton\per\metre}$) indicating that mmA is much larger than the actual size of the particles.
Upon decreasing the area to mmA=\SI{110}{\nano\metre\squared}, the surface pressure increases with a linear slope.
The resistance of the film upon compression increases indicated by the higher modulus E ($15 - \SI{23}{\milli\newton\per\metre}$).
Further compression  leads to a shoulder in the isotherm at $\text{mmA}\approx \SI{40}{\nano\metre\squared}$ accompanied by a sharp maximum in E.

The particle film is resisting further compression before structural changes in the film lead to a release of the stress. The two dashed black lines in the left panel of figure \ref{fig:isotherme_fenp5} denote the minimum and maximum projected area of the spheroidal particles based on their size as determined by SAXS. Here we take an oleic shell thickness of \SI{1.25}{\nano\meter} into account, thus allowing for interpenetration of shells. Even for this densely packed configuration the minimum projected area is significantly larger than the position of the observed shoulder in the isotherm. This points to an inappropriate determination of the mmA. A possible reason is an incorrect value of the amount of particles spread on the surface. We will discuss this in detail later on when looking at the XRR profiles.

The observed shoulder in the isotherm may be explained by either a collective flipping of the ellipsoidal particles, or by buckling of the film and multilayer formation. A collective flipping process seems to be very unlikely, instead single flips due to a locally increased density in the monolayer are expected.\cite{Basavaraj2006} Thus, we suppose that buckling of the film is responsible for the observed feature in the isotherm, which will be further investigated by in-situ X-ray reflectivity and discussed in the following paragraph.

The right panel of figure \ref{fig:isotherme_fenp5} shows the isotherm and elastic modulus of FeNP10 films.
The films were also prepared for low surface coverage ($\Pi \approx \SI{1}{\milli\newton\per\metre}$), however the pressure immediately started to increase with compression of the film.
From $\text{mmA} = \SI{300}{\nano\metre\squared}$ to \SI{100}{\nano\metre\squared}, the surface pressure increases with a linear slope.
In this region the resistance of the film, characterised by the elastic modulus (E ranging between $12 - \SI{20}{\milli\newton\per\metre}$), is similar to the smaller particles at the same surface pressure.
Afterwards, the slope of the isotherm starts to change and increases more strongly. As already seen for the FeNP5 system, the projected particle area deduced from SAXS results is much larger than this feature of the isotherm. This discrepancy will be further discussed in the following section. In contrast to the FeNP5 film no shoulder in the isotherm is observed.

The change in the slope of the isotherm is accompanied by an increase of E up to $\SI{75}{\milli\newton\per\metre}$.
The magnitude of the modulus helps to identify the structural state of the two particle films.
Elastic moduli for nanoparticle films have been observed to range between \SI{10}{\milli\newton\per\metre} and up to several \SI{100}{\milli\newton\per\metre}.\cite{Choudhary2015,Vegso2012}
The maximum modulus depends on shell ligand, size distribution and spacing between the particles.
Our samples have a maximum of $\text{E} \approx \SI{48}{\milli\newton\per\metre}$ for FeNP5 and $\text{E} \approx \SI{75}{\milli\newton\per\metre}$ for FeNP10 which is in agreement with published data on iron oxide nanoparticles.\cite{You2013}
Both values agree with the modulus characteristic for an oleic acid monolayer in the liquid-expanded state.\cite{Baba2013}
In case of the FeNP5 films, the value is considerably lower because the isotherm indicates that structural changes occur in the film before the modulus increases further.
The sharp increase can be understood as resistance of the oleic acid ligand because further interdigitation of the ligand is restricted by its double bond that reduces their ability to interpenetrate.
We interpret the lower value of $E \approx 12-\SI{23}{\milli\newton\per\metre}$ in the first part of both isotherms as a resistance of rather loosely connected groups of particles that slide past each other while the maximum represents real compression of the shell in a dense layer.

\begin{figure}[h!]
	\centering
	\includegraphics[width=0.49\textwidth]{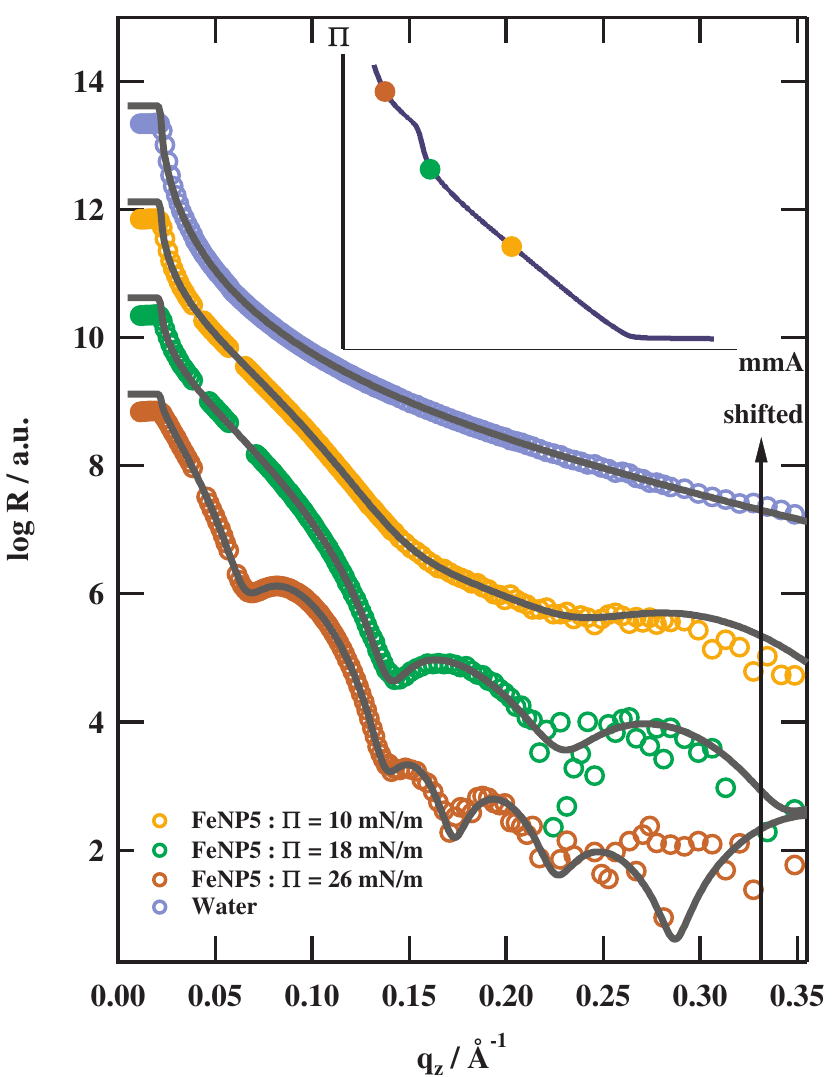}
	\includegraphics[width=0.49\textwidth]{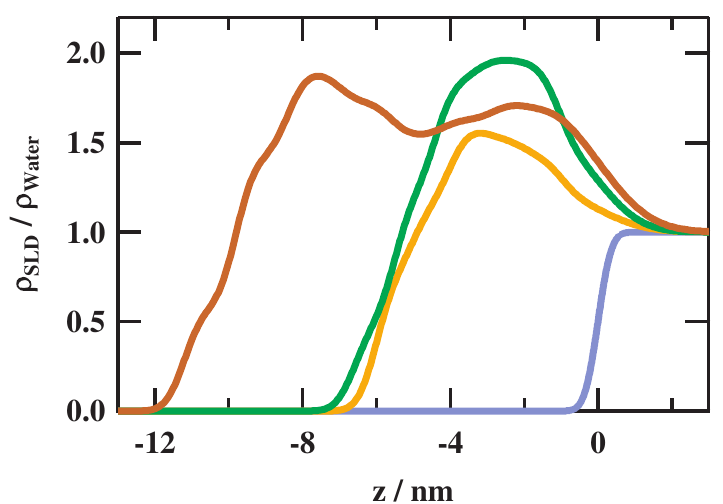}
	\begin{small}
	
	\vspace{2mm}
	\begin{tabular}{l c c c c c}
	\hline
	$\Pi$ & number of boxes $N$ & box size $a$ & width $\sigma$ \\ \hline
	\SI{10}{\milli\newton\per\metre}  & 6 & \SI{1.0}{\nano\metre} & \SI{0.4}{\nano\metre} \\ 
	\SI{18}{\milli\newton\per\metre}& 6 & \SI{1.1}{\nano\metre} &  \SI{0.4}{\nano\metre} \\ 
	\SI{26}{\milli\newton\per\metre}& 8 & \SI{1.4}{\nano\metre} &  \SI{0.4}{\nano\metre} \\ \hline
	\end{tabular}
	\end{small}
	\caption{X-ray reflectivity data of FeNP5 films measured at three different positions in the isotherm. For comparison the reflectivity of the bare water interface was measured prior to film preparation. The solid lines demonstrate the good agreement between the used box-model (equation \ref{eq:multilayer_model}) and the data. The resulting normalised scattering length density profiles are shown below. The water surface is placed at $z=0$ and negative values of $z$ correspond to increasing height above the water surface. Important fitting parameters are depicted in the table below.}
	\label{fig:fenp5}
\end{figure}

\subsection{Vertical Density Profiles of Nanoparticle Layers at the Air-Water Interface} 
\begin{figure}[t!]
	\centering
	\includegraphics[width=0.49\textwidth]{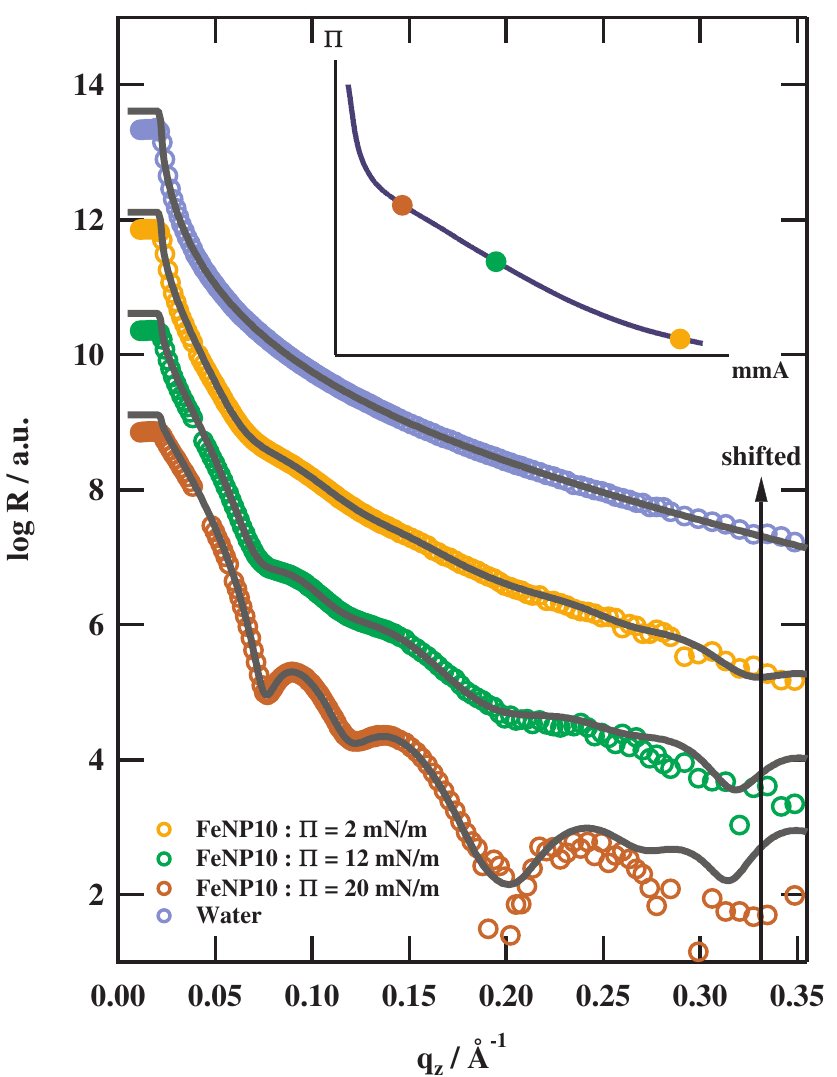}
	\includegraphics[width=0.49\textwidth]{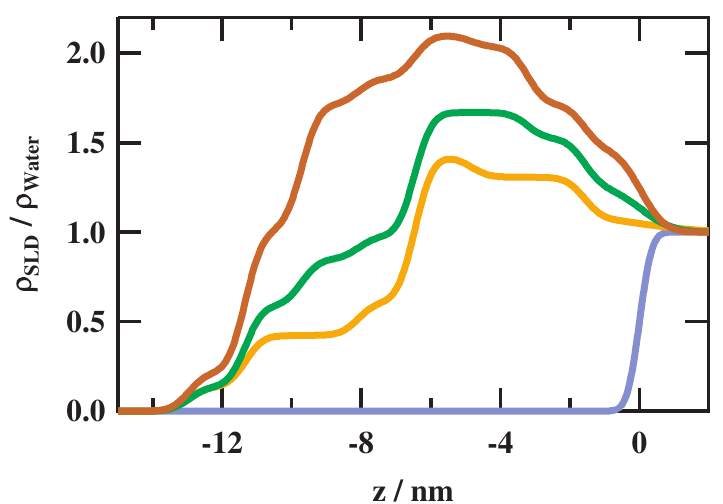}
	\begin{small}
	
	\vspace{2mm}
	\begin{tabular}{l c c c c c}
	\hline
	$\Pi$ & number of boxes $N$ & box size $a$ & width $\sigma$ \\ \hline
	\SI{2}{\milli\newton\per\metre}  & 8 & \SI{1.65}{\nano\metre} & \SI{0.4}{\nano\metre} \\ 
	\SI{12}{\milli\newton\per\metre} & 8 & \SI{1.65}{\nano\metre} &  \SI{0.4}{\nano\metre} \\ 
	\SI{20}{\milli\newton\per\metre} & 8 & \SI{1.65}{\nano\metre}  &  \SI{0.4}{\nano\metre} \\ \hline
	\end{tabular}
	\end{small}
	\caption{X-ray reflectivity data of FeNP10 films measured at three different positions in the isotherm. For comparison the reflectivity of the bare water interface was measured prior to film preparation. The solid lines demonstrate the good agreement between the used box-model (equation \ref{eq:multilayer_model}) and the data. The resulting normalised scattering length density profiles are shown below. The water surface is placed at $z=0$ and negative values of $z$ correspond to increasing height above the water surface. Important fitting parameters are depicted in the table below.}
	\label{fig:fenp10}
\end{figure}

Film thickness was determined by specular reflectivity measurements for several positions in the isotherm of both particle sizes.
Due to the high scattering contrast of the particles, an interference pattern can be clearly identified in the raw data (see figures \ref{fig:fenp5} and \ref{fig:fenp10}) compared to the reflectivity of the bare water interface.
When surface pressure is increased, the number of minima in the reflectivity increases.
They become more distinct and shift to smaller $q$ at high surface pressure.   
These patterns confirm the formation of a particle film which becomes denser upon compression.
The missing data points in the raw data for small $q \leq \SI[per-mode=reciprocal]{0.05}{\per\angstrom}$ are caused by absorber switches.
Due to the high scattering contrast, intensity was too high after the absorber switch, thus not properly recorded by the detector.

The scattering length density perpendicular to the interface is described by $N$ equally sized boxes. 
In each box of size $a$, the scattering length density is constant.
The edge of each box is modulated by an error function with a width $\sigma$ to smooth the transition between two neighbouring boxes.
To keep the model as simple as possible, the lowest number of boxes $N$ resulting in a reasonable description of the data was chosen.
Finally, the scattering length density profile of the film is given by
\begin{equation}\label{eq:multilayer_model}
	\langle \rho(z) \rangle = \sum_{j=1}^N \frac{\Delta \rho_j}{2}\left(1+\text{erf}\left(\frac{z+j\cdot a}{\sqrt{2}\sigma}\right)\right) + \frac{\rho_{\text{water}}-\sum_{j=1}^N\Delta \rho_j}{2}\left(1+\text{erf}\left(\frac{z}{\sqrt{2}\sigma_0}\right)\right)
	\end{equation}
with $\Delta \rho_j$ as the increase or decrease of the scattering length density between each box and $z$ as the distance from the air-water interface ($z<0$ for air).

\begin{figure}[t]
	\centering
	\includegraphics[width=0.8\textwidth]{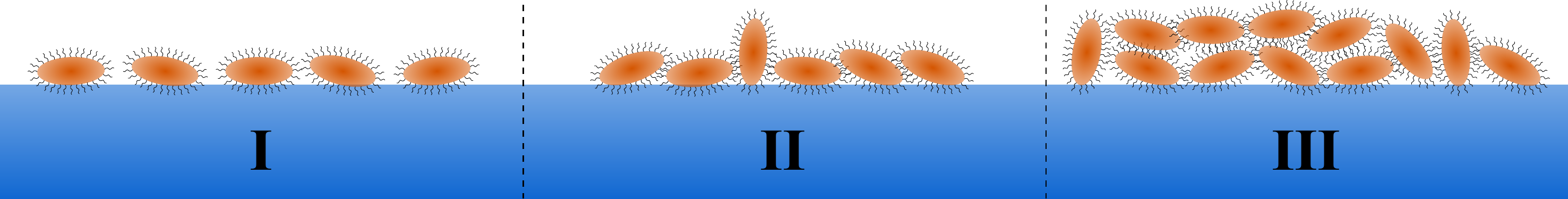} 
	\caption{The sketch illustrates the possible different orientations of anisotropic FeNP5 particles at the air-water interface. The three regimes correspond to the three different surface pressures investigated by XRR.}
	\label{fig:fenp5_sketch} 
\end{figure}

Film thickness was measured for three different positions in the isotherm of the FeNP5 film.
The raw reflectivity curves are shown in figure \ref{fig:fenp5} compared to the bare water interface.
Their positions in the isotherm are highlighted in the inset.
In all curves a very good agreement between model and experimental data is found as demonstrated by the full black lines in the figure.
Scattering length density profiles normalised by $\rho_{\text{water}}$ are shown in the right panel of the figure.
The other important parameters of the model ($N$, $a$ and $\sigma$) are summarized in the table below the figure.
For a surface pressure of $\Pi =\SI{10}{\milli\newton\per\metre}$ (yellow points), a broad minimum can be observed in raw data.
A model for a single particle film of \SI{6}{\nano\metre} thickness is able to describe the data.
The second measurement was performed at $\Pi = \SI{18}{\milli\newton\per\metre}$ (green points) just before the transition in the compression isotherm occurs.
The minimum separates into two clear minima and the model for a single particle film still agrees with the experimental data.
The thickness slightly increases to \SI{6.6}{\nano\metre} while the scattering length density profile also indicates that the particle film is more compressed.
The last measurement was performed at $\Pi = \SI{26}{\milli\newton\per\metre}$ right after the transition occurred.
The appearance of two additional minima already indicates that the structure of the film changed substantially.
A model consisting of $8$ boxes provides a very good description of the data for q$_{\text{z}} \leq \SI[per-mode=reciprocal]{0.25}{\per\angstrom}$.
The overall thickness of the layer increases to \SI{11.2}{\nano\metre}.
The scattering length density profile also changes its shape.
Instead of one single maximum positioned roughly at half the particle size, a second maximum at larger distances from the interface appears. From the shape of the isotherms we already expected multilayer formation when increasing the surface pressure beyond \SI{18}{\milli\newton\per\metre}. This picture is confirmed by the SLD profile measured at \SI{26}{\milli\newton\per\metre}. The overall film thickness almost doubles and the appearance of the second maximum points to the formation of a second layer of nanoparticles on top of the already existing layer. A cartoon-like representation of the arrangement of nanoparticles with increasing pressure is shown in figure \ref{fig:fenp5_sketch}.

Furthermore, the absolute values of the scattering length density curves allow to calculate the particle density in the film. For bulk iron oxide, the scattering length density is $\rho_\text{iron oxide}=\SI[per-mode=reciprocal]{40.7e-6}{\per\angstrom\squared}$, four times larger than water ($\rho_\text{water}=\SI[per-mode=reciprocal]{9.4e-6}{\per\angstrom\squared}$). Since the oleic acid shell does not show any scattering contrast to water, we estimate the scattering length density of the shell to be equal to that of water. Assuming interpenetrating shell ligands, as already discussed above, and a maximum packing density of two dimensional spheres or ellipses of 0.9, the maximum scattering length density of the film can be calculated to $\rho_{\text{SLD}_{min}}=2.31\cdot\rho_\text{water}$ and $\rho_{\text{SLD}_{maj}}=2.83\cdot\rho_\text{water}$ for standing and flat configuration respectively. From the shape of the isotherm discussed in accordance to figure \ref{fig:isotherme_fenp5}, we would expect a dense packing of the flat configuration right before the shoulder, around the $\Pi=\SI{18}{\milli\newton\per\metre}$ measurement. The maximum value of the scattering length density profile in this case is $1.9\cdot\rho_\text{water}$, roughly \SI{70}{\percent} of the estimated value. Thus, the particle density in the film has to be reduced accordingly, which means that the mean molecular areas discussed in the previous paragraph seem strongly overestimated (roughly by a factor of three based on the XRR profiles). These interpretations are further supported by observations during film preparation. After spreading the sample solution a substantial staining of the metal parts of the Hamilton syringe is visible, a clear indication of remaining nanoparticles that are missing in the layer.

For the FeNP10 film, reflectivity was also measured for three different positions in the isotherm.
Raw data is shown in figure \ref{fig:fenp10} with the measurement position highlighted in the inset of the figure.
Interference patterns are visible for all film measurements.
The most distinct minima are visible for the highest surface pressure, however, the same features are also visible for the two lower surface pressures.
There is no strong shift in the positions.
A very good agreement between model and experimental data is demonstrated by the full lines for $\text{q}_{\text{z}} \leq \SI[per-mode=reciprocal]{0.25}{\per\angstrom}$.
The model described by equation (\ref{eq:multilayer_model}) is used with 8 boxes, each having a size of \SI{1.65}{\nano\metre} and a $\sigma$ of \SI{0.4}{\nano\metre}.
The overall thickness of \SI{13.2}{\nano\metre} for the layer compares very well with the particle size obtained from the form factor fit of the SAXS data.

The normalized scattering length density profiles at the right panel of figure \ref{fig:fenp10} indicate that particles are moving closer together upon compression. Scattering length density increases but film thickness is the same.  We see no indication for multilayer formation which is consistent with results from Langmuir isotherms.

Analysing the absolute values of the scattering length density, the same overestimation of mmA as discussed for the FeNP5 particles is found. For a densely packed film of FeNP10 particles with interpenetrating shell a maximum scattering length density of $3.0\cdot\rho_\text{water}$ can be estimated. From the shape of the isotherm such a dense packing can be assumed at the highest investigated surface pressure of $\Pi=\SI{20}{\milli\newton\per\metre}$. However, we only measure a value of $2.1\cdot\rho_\text{water}$, which is \SI{70}{\percent} of the expected value. Again, the mmA seems to be overestimated by a factor of three as already discussed for the smaller particles.

\section{Polymer Nanocomposite Films}
The two presented sizes of iron oxide nanoparticles form single particle layers at the air-water interface (at sufficiently low surface pressure) and therefore seem to be promising candidates for the formation of polymer nanocomposite films.
In a previous publication, we presented a block copolymer PEG$_6$-\textit{b}-P\textit{n}BA$_{132}$ that forms very stable films at the air-water interface.\cite{Appel2018}
The polymer spreads as a thin layer (thickness around $2-\SI{3}{\nano\metre}$) and has a very characteristic isotherm.
One unique feature is a long constant pressure plateau in the isotherm in which polymer chains dewet from the air-water interface.
Furthermore, the 2D radius of gyration for one chain is approximately \SI{60}{\nano\metre\squared} which is comparable to the size of the FeNP5 and FeNP10 particles (FeNP5: 27 or \SI{70}{\nano\metre\squared} and FeNP10: \SI{129}{\nano\metre\squared}).

The two composite solutions were prepared by adding polymer to the  iron oxide nanoparticles dissolved in chloroform.
SAXS measurements of the form factor (data shown in ESI) demonstrate that addition of polymer does not lead to aggregation of the particles. The same particle form factors are observed as in the toluene solutions of the nanoparticles.
The particle concentration for both composites was chosen such as to have a coverage of $\SI{3}{\percent}$ of the area in a film prepared at the air-water interface before compression.
The composite films were prepared for low surface coverage and compression isotherms were recorded.
The mixture of FeNP5 will be referred to as NPC5, and NPC10 for FeNP10.

\subsection{Langmuir Isotherms of Composite Layers}
\begin{figure}[t]
	\centering
	\includegraphics[width=0.49\textwidth]{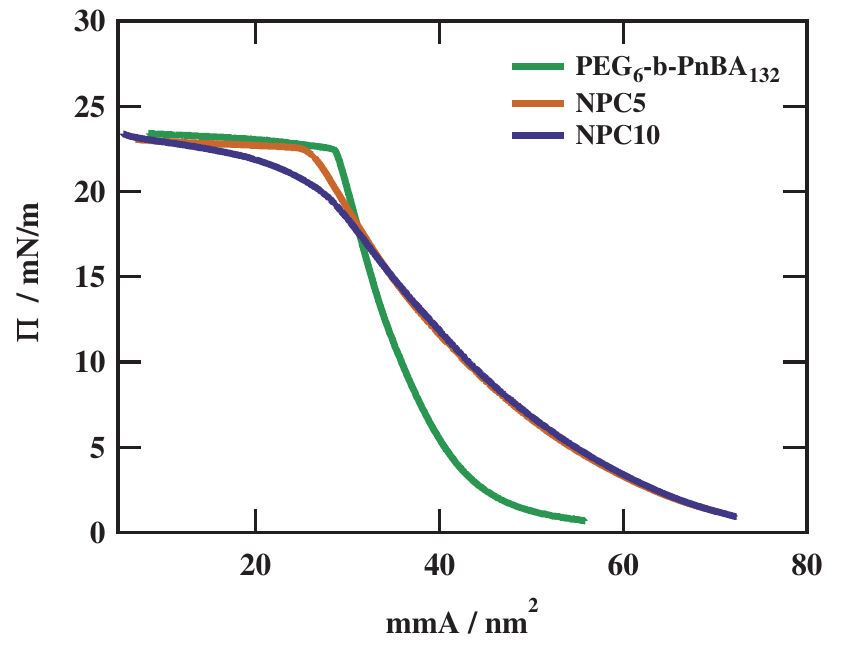}
	\includegraphics[width=0.49\textwidth]{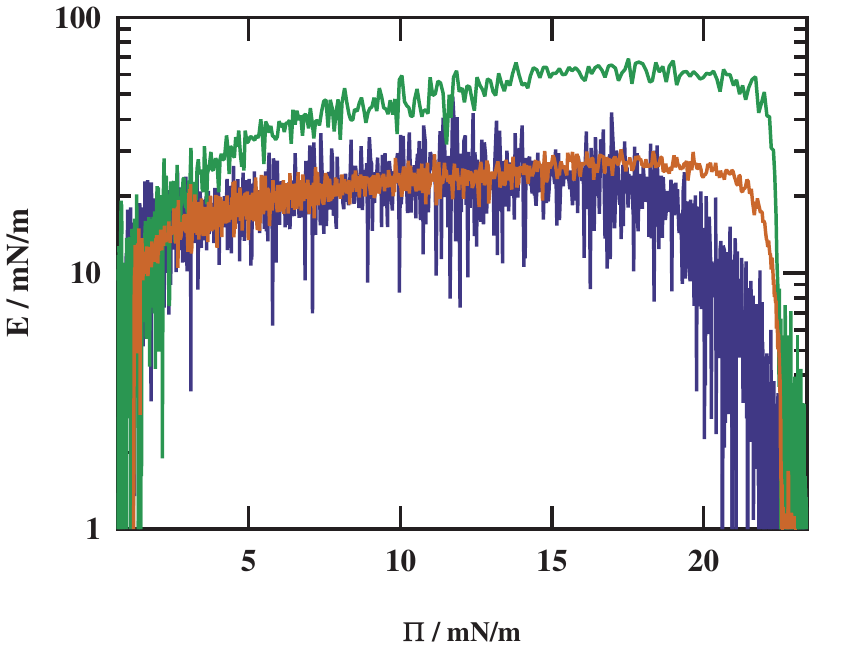}
	\caption{Langmuir isotherms presented as surface pressure ($\Pi$) vs. mean molecular area (mmA) for NPC5, NPC10 and the pure polymer film. The elastic modulus ($E$) computed from the isotherms is depicted below as $E$ vs. $\Pi$.}
	\label{fig:isotherme_npc}
\end{figure}
We now compare the compression isotherms of the composites with the isotherms of the pure polymer layer.
Results of these measurements  are shown in figure \ref{fig:isotherme_npc}.
They are presented as surface pressure $\Pi$ over the mean molecular area mmA for a polymer chain.
For the composites, the area covered by nanoparticles is subtracted from the mmA so that the composite and polymer isotherms can be directly compared.
The isotherms of NPC5 and NPC10 are closer to the polymer isotherm than to that of the pure nanoparticles  (see figure \ref{fig:isotherme_fenp5}).
The surface pressure increases for similar areas indicating that the polymer is predominately responsible for the compression behaviour of the composite.
However, in detail the compression behaviour of the three systems is different.
For the composite films the surface pressure starts to increase already for larger areas compared to the polymer film.
After reaching a critical surface pressure of $\Pi_{\text{C}} \approx 22-\SI{23}{\milli\newton\per\metre}$, a kink can be observed for the pure polymer and NPC5 film at almost the same value of mmA. Further compression of both films does not lead to any increase in surface pressure. In case of the pure polymer system, the constant pressure plateau has been identified as polymer dewetting the air-water interface.\cite{Appel2018} The overall similarity between the pure polymer and NPC5 isotherms suggests that this dewetting mechanism is also found in the composite. No signs of buckling and multilayer formation are found is this isotherm. This is presumably due to the significantly reduced density of nanoparticles in the composite film, in comparison to the pure nanoparticle ones.

For NPC10 the similarity to the pure polymer isotherm is not as obvious. Although the low pressure regime of the composite isotherms ($\Pi \leq \SI{15}{\milli\newton\per\metre}$) is almost identical, neither a sharp kink nor a constant pressure plateau can be observed for the NPC10 system. Instead the slope of the NPC10 isotherm starts to flatten out already at a higher mmA but without reaching a plateau at the highest investigated compressions.

The right panel of figure \ref{fig:isotherme_npc} shows the elastic modulus $E$ vs. $\Pi$ as a representative measure of the resistance of the film upon exhibiting an external force (see equation (\ref{eq:compressibility})).
In case of the polymer film, a broad plateau can be observed with $E_{\text{max}} \approx \SI{60}{\milli\newton\per\metre}$ before asymptotically vanishing for the critical surface pressure $\Pi_{\text{C}}$.
For NPC5, the shape of $E$ is quite similar.
However, absolute values of the modulus are different.
In case of the pure FeNP5 film, we measure a lower $E$ modulus (see figure \ref{fig:isotherme_fenp5}) which is caused by the interactions between the oleic acid shell molecules.
The $E$ modulus of the composites is reduced by the interaction of shell molecules with surrounding particles and polymers.
The fluctuations of $E$ are stronger for the composite compared to the polymer while we observe the same deviation upon reaching $\Pi_{\text{C}}$.
When comparing NPC10 to NPC5, two disparities become apparent.
First of all, the $E$ modulus decreases more slowly for higher surface pressures ($\Pi >\SI{15}{\milli\newton\per\metre}$).
The absolute value for $E$ is approximately the same, but the fluctuations become even more pronounced.
These fluctuation may indicate the existence of larger structures within the film.
Obviously these structures have to be related to the iron oxide nanoparticles and the fluctuations may be related to aggregation of nanoparticles in the polymer matrix.
FeNP10 affects the elastic properties of the film more strongly than FeNP5 although the area covered by particles is the same for both composites.
To further investigate this behaviour it is therefore important to obtain information on the distribution of the nanoparticles inside the polymer matrix.
To find this distribution we followed two different experimental approaches.
The first one was to determine the film thickness and density profile by performing in-situ XRR experiments on the composites spread at the air-water interface.
In another experiment, the composites were transferred on silicon wafers using the Langmuir-Blodgett technique and an ex-situ AFM experiment was used to investigate the distribution of particles.
Results of the in-situ approach will be discussed first. 

\subsection{Vertical Density Profiles of Composite Layers at the Air-Water Interface}
\begin{figure}[!t]
	\centering
	\includegraphics[width=0.49\textwidth]{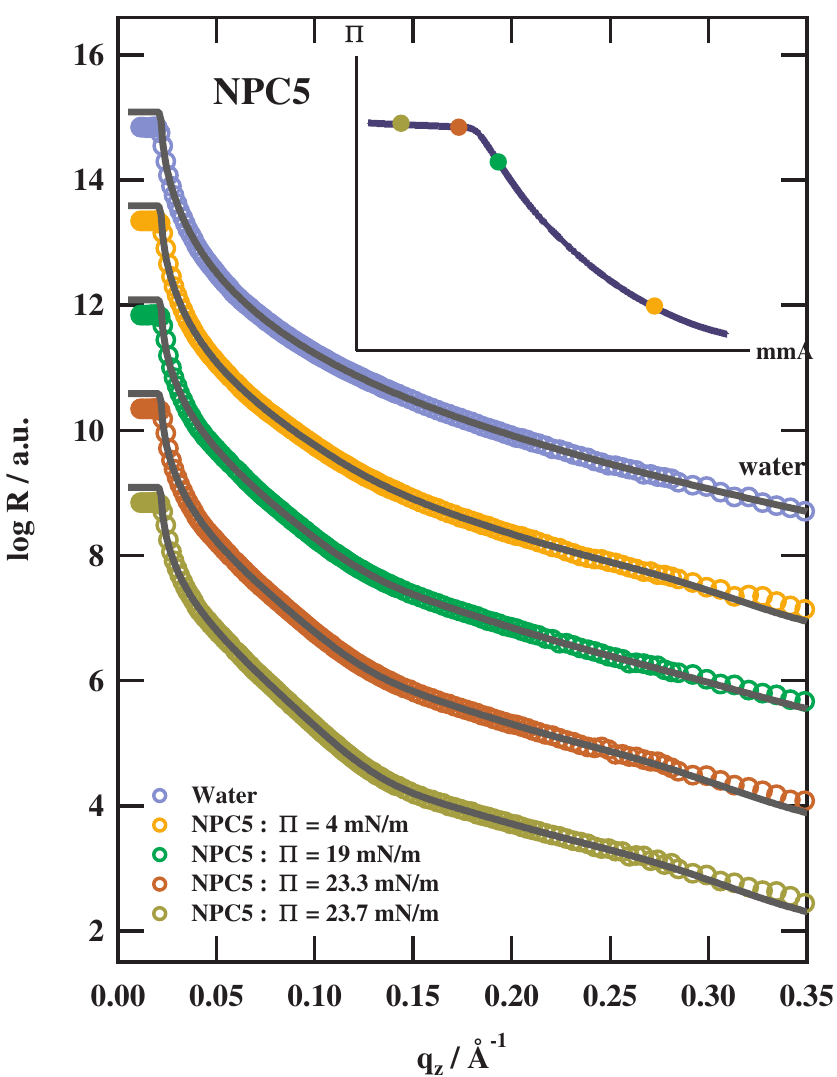}
	\includegraphics[width=0.49\textwidth]{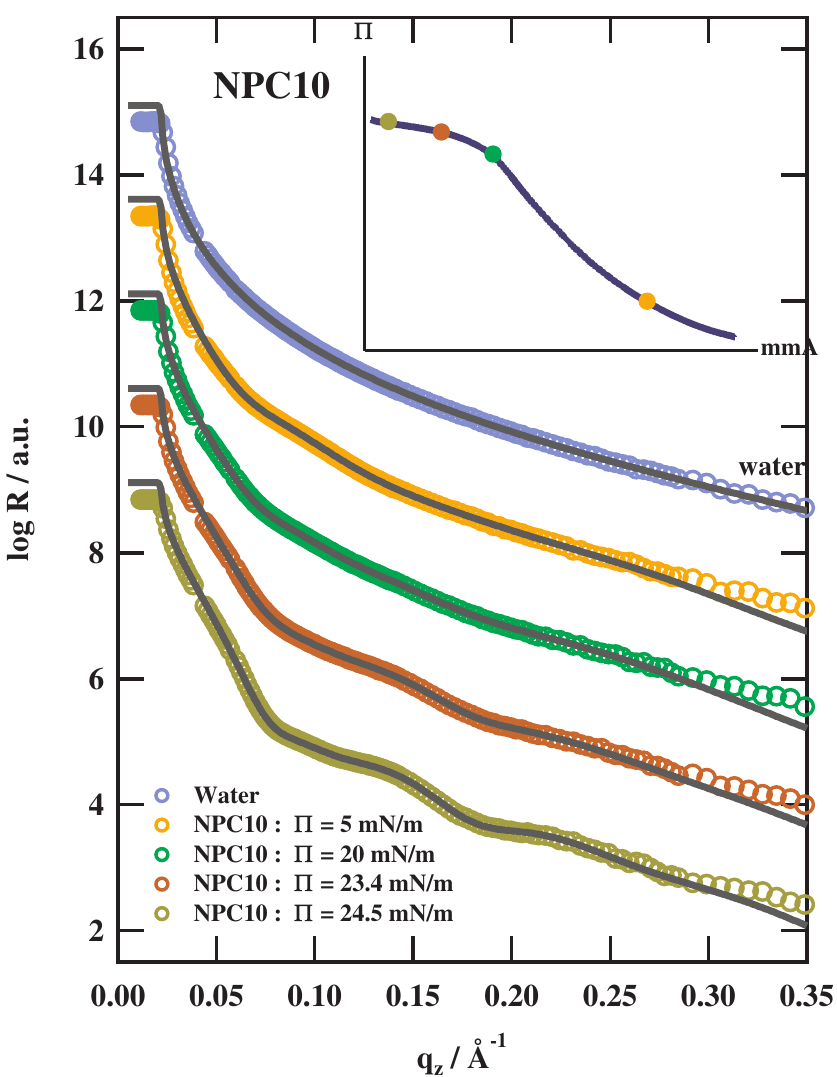}
	\includegraphics[width=0.49\textwidth]{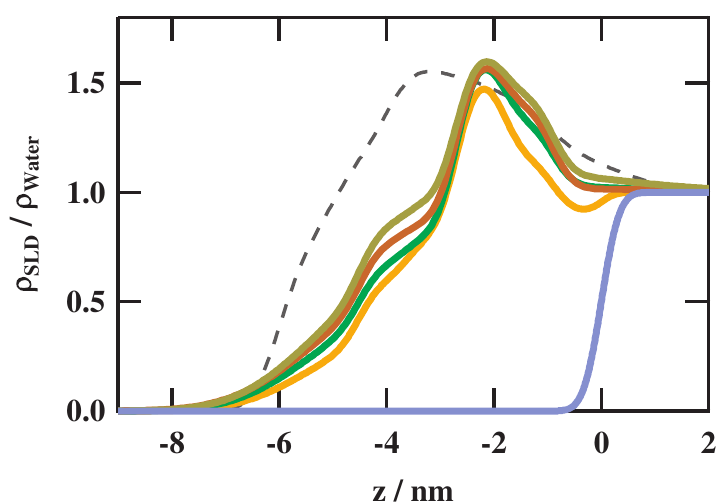}
	\includegraphics[width=0.49\textwidth]{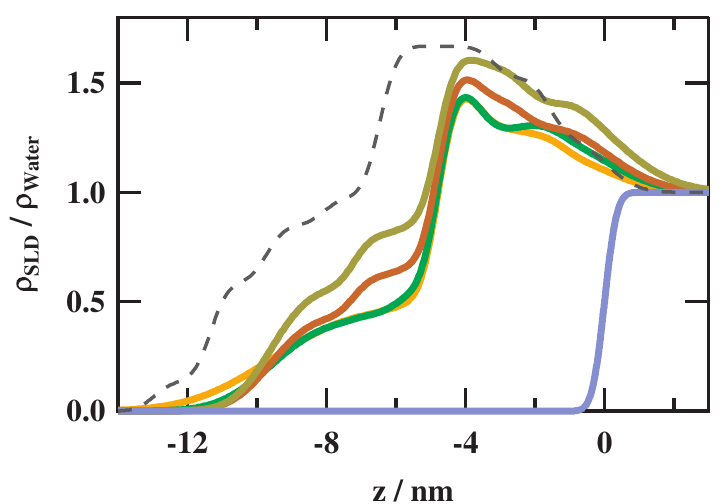}
	\caption{X-ray reflectivity data of NPC5 and NPC10 films both measured at four different positions in the isotherm. For comparison the reflectivity of the bare water interface was measured prior to film preparation. The solid lines demonstrate the good agreement between the used box-model (equation \ref{eq:multilayer_model}) and the data. The resulting normalised scattering length density profiles are shown below. The dashed lines represent the SLD profile at intermediate pressure of the respective pure particle film (details see text).}
	\label{fig:npc_XRR} 
\end{figure}
Specular reflectivities were measured in-situ for four positions in the isotherms of both composite films to study film morphology.
Results for both nanoparticles and films at different compressions are compiled in figure \ref{fig:npc_XRR}.
Due to the high contrast of the iron oxide nanoparticles and the smoothness of the films, interference patterns can be observed in the raw data of both composite films (upper panel of figure \ref{fig:npc_XRR}).
Up to a scattering vector of $\text{q}_{\text{z}} \leq \SI[per-mode=reciprocal]{0.3}{\per\angstrom}$ a very good agreement between the raw data and the multilayer model introduced in equation (\ref{eq:multilayer_model}) is demonstrated by the full black lines.
Six boxes of equal size ($a=\SI{0.9}{\nano\metre}$ and $\sigma = \SI{0.3}{\nano\metre}$) were used to describe the NPC5 film, while eight were necessary for NPC10 ($a=\SI{1.2}{\nano\metre}$ and $\sigma = \SI{0.4}{\nano\metre}$).
This leads to an overall thickness of \SI{5.6}{\nano\metre} for NPC5 and \SI{9.6}{\nano\metre} for NPC10.
This shows that no multilayer is formed for either size of nanoparticles, instead particles are immersed in the polymer layer.
The normalized scattering length density profiles in the bottom panel of figure \ref{fig:npc_XRR} are shown for all investigated positions in the isotherm and compared to that of the pure nanoparticle films (dashed grey lines, measured at $\Pi = \SI{10}{\milli\newton\per\metre}$ for FeNP5 and $\Pi = \SI{12}{\milli\newton\per\metre}$ for FeNP10).

The maximum scattering length density is almost equal for composites and pure nanoparticle films (intermediate surface pressure).
However, the shape of the composite profiles have one common feature that clearly 
separates them from the pure particle films, which is the sharp decrease of the scattering length density for $z \approx -\SI{3}{\nano\metre}$ (NPC5) respectively $z \approx -\SI{4}{\nano\metre}$ (NPC10). This feature can be related to the upper limit of the polymer film.
The scattering length density of the polymer is similar to water or the shell of the particles.
The reflectivity experiment measures a laterally averaged scattering length density. This average comprises nanoparticles and polymer for small distances from the substrate and nanoparticles and air for distances larger than the polymer film thickness.
Therefore, we can identify the polymer film in the density profiles of the composite films with a thickness of \SI{3}{\nano\metre} to \SI{4}{\nano\metre}. This is in good agreement with the thickness of the pure polymer film.\cite{Appel2018}

The in-situ XRR measurements show that nanoparticles are present within the composite films. 
However, specular reflectivity measures the film thickness as an average over the interface, in contrast to that off-specular scattering gives access to the lateral distribution of the particles inside the film.
Off-specular data was also recorded for the films, but it was not possible to separate the scattering of a single particle (form factor) from the contribution of the height-height correlation function (see ESI).
Still, the different compression behaviour for the two composites indicates that there must be a structural difference. In order to follow this question further LB films were transferred on silicon wafers at $\Pi = \SI{15}{\milli\newton\per\metre}$ to be investigated  using AFM. This should provide information on the lateral distribution of nanoparticles in the composite layer.

\subsection{Lateral Distribution of Nanoparticles in Composite Layers}
\begin{figure}[!t]
	\centering
	\hspace{3mm}
	\includegraphics[width=0.48\textwidth]{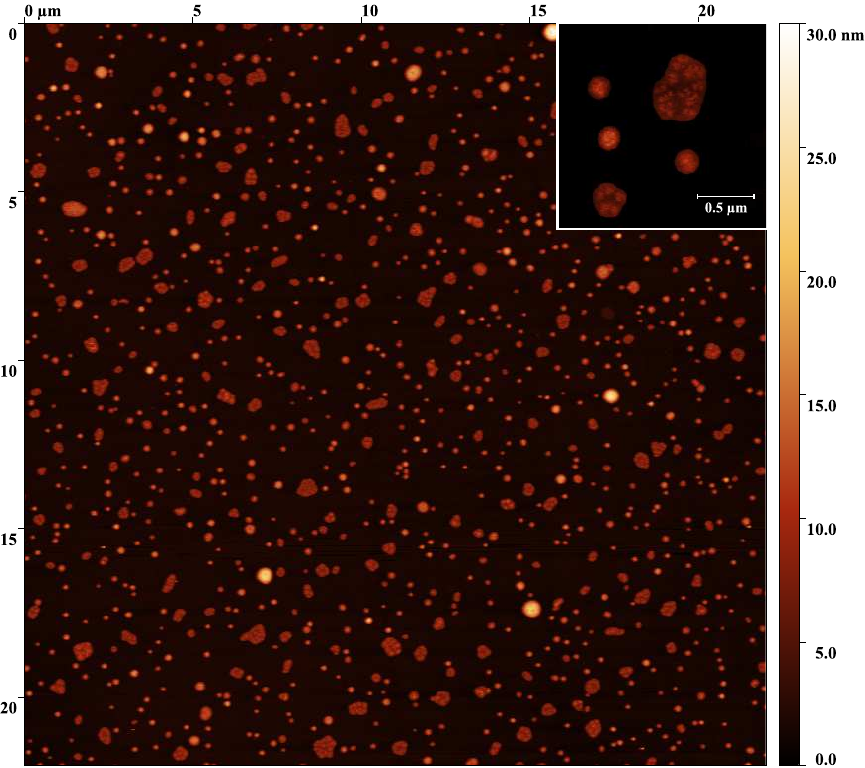}
	\hfill
	\includegraphics[width=0.48\textwidth]{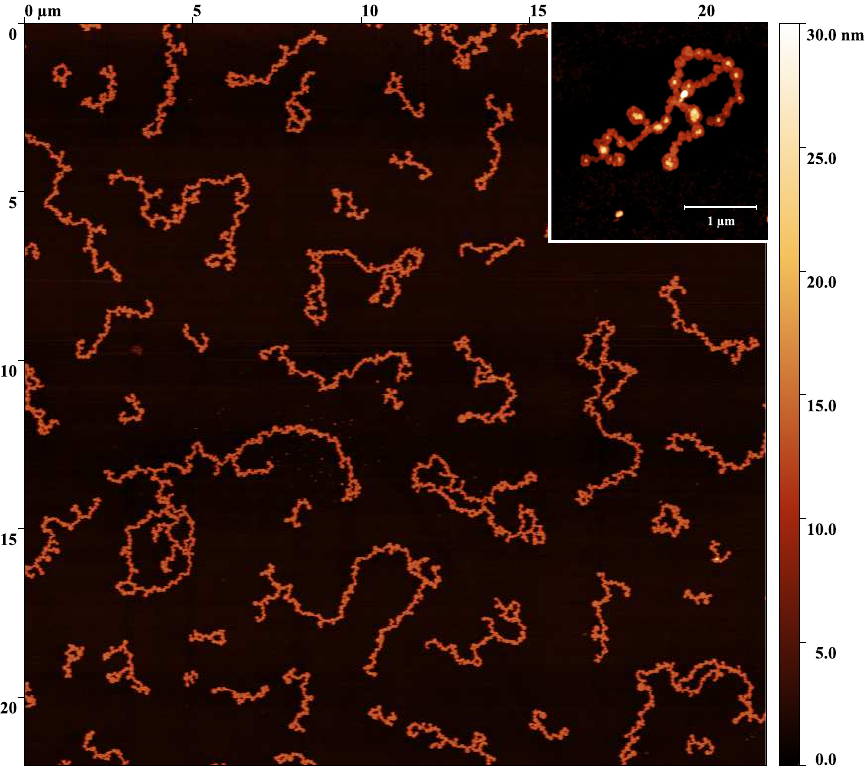}
	
	\vspace{7mm}
	\includegraphics[width=0.46\textwidth]{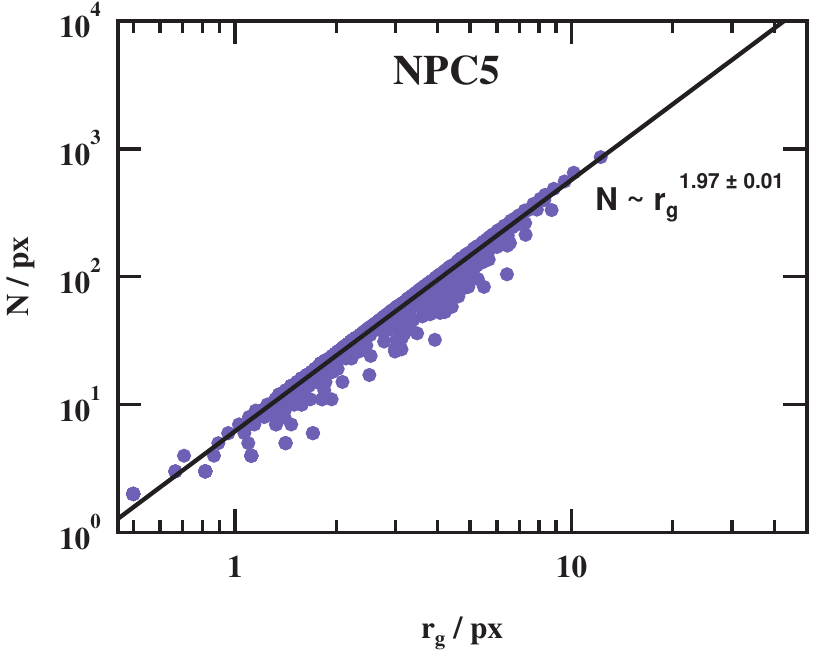}
	\includegraphics[width=0.46\textwidth]{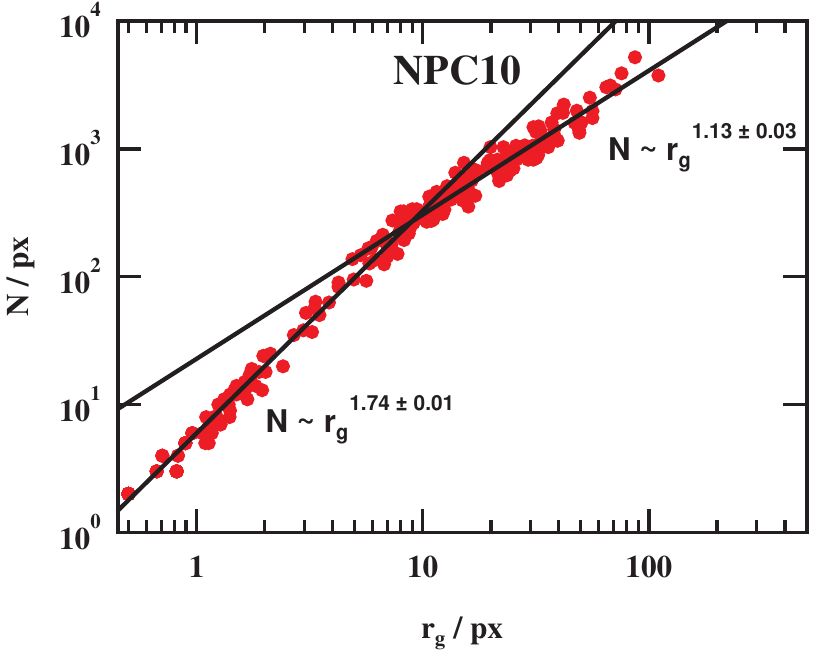}
	\caption{Representative atomic force microscopy images (\SI{22}{\micro\meter} x \SI{22}{\micro\meter}) for NPC5 and NPC10 LB films. High resolution scans of characteristic domains are shown as inset. The bottom panel shows a fractal analysis in which images from three different regions were used for both systems. For every identified object its area (given by the respective number of pixels $N$) is plotted versus its radius of gyration $r_g$. The solid lines are power-law fits to the data.}
	\label{fig:AFM_NPC}
\end{figure}

Results of the AFM measurements are shown in figure \ref{fig:AFM_NPC} for NPC5 (left panel) and NPC10 (right panel).
Both images show an area of $\SI{22}{\micro\metre}$ x $\SI{22}{\micro\metre}$ while the maximum height of the observed structures is less than $\SI{30}{\nano\metre}$.
Therefore, particles are essentially confined in  2D.
The observed objects are distributed in a continuous matrix, which has to be the polymer film. We thus have a complete coverage of the investigated area on the silicon waver by the polymer nanocomposite.
We note that lateral dimensions are a convolution of cantilever and object size, thus lateral sizes appear significantly larger.\cite{Sievers2012}
The insets in the images show higher resolution scans of selected regions on the silicon wafer. 
Single nanoparticles can be clearly identified as shown in the inset of the figure and there is a very obvious difference between NPC5 and NPC10 films.
In case of NPC5, the lateral distribution of FeNP5 in the polymer matrix is given by many almost disk-like domains (left panel). The high resolution image indicates the presence of single particles as well as small clusters of only few particles.
Obviously, the aggregation behaviour of both nanoparticles is very different.
For NPC10, we observe chain-like structures of different sizes (right panel).
To quantitatively investigate the different aggregation behaviour of NPC5 and NPC10, we analyse these images further. 
We first identify each connected domain with a height above \SI{4}{\nano\metre} as one single object. For every object the radius of gyration $\text{r}_{\text{g}}$ in pixels ($\text{1 px}\  \widehat{=}\ 30\,\mathrm{nm}$) is calculated and compared to its area (defined by the number of occupied pixels N). Images from three different positions of the LB film were used to improve statistics.
In total 4644 objects for NPC5 and 266 for NPC10 were considered.

In the bottom of figure \ref{fig:AFM_NPC}, we present the results from these considerations.
There is a power law relation between the number N and $\text{r}_{\text{g}}$ ($\text{N} \sim {\text{r}_{\text{g}}}^d$) with a fractal dimension $d$.
We identify a single power law for NPC5 for the full range of observed aggregate sizes while there is a crossover between two different exponents for NPC10.
From the exponent of the power law we can determine the fractal dimension $d$ of the structures.
The upper limit for 2D structures is $d=2$ (compact object) and $d=1$ is the lower limit (straight line).
We note that both limits are logical boundaries, however the upper one can be seen in the data on NPC5.

The fractal dimension of $d_{\,\text{NPC5}} = 1.97 \pm 0.01$ confirms our initial description of almost disk-like objects for FeNP5.
In case of NPC10, we find two different power laws.
For large structures, the fractal dimension is $d_{\,\text{NPC10,large}} = 1.13 \pm 0.03$ which is close to the limit of a straight line.
This agrees with our initial interpretation of chain-like structures of FeNP10.
However, it is interesting that we observe a crossover of the power law for small domains ($\text{r}_{\text{g}} < 9\,\mathrm{px}$).
Small domains tend to be less extended with a fractal dimension of $d_{\,\text{NPC10,small}} = 1.74 \pm 0.01$.

Particles with dominating magnetic interaction are expected to form chain like structures, with dipole moments aligned parallel along the chain.
Although predicted theoretically and found in many simulations, isolated chains in zero field are rarely observed experimentally.\cite{deGennes1970,Pastor1995,Holm2005,Prokopieva2009,Lefebure1998,Vorobiev2015} 
So far, clear evidence has only been obtained by vitrified cryo-TEM samples.\citep{Butter2003,Klokkenburg2006} 
For a dried powder of FeNP10 we can clearly measure a remanent magnetisation $M_\text{R} =\SI{0.019}{\ampere\nano\metre\squared}$ per particle (see ESI). Two-dimensional simulations of aggregating magnetic particles indeed find a fractal dimension of $d = 1.13 \pm 0.01$ for large cluster sizes (several thousand particles).\cite{Pastor1995}
This result nicely agrees with our finding for large cluster sizes even though k$_{\text{B}}$T should be much larger than the magnetic dipolar interaction energy between two single particles.
Here, the presence of the polymer matrix turns out to be the important factor.

A particle approaching a large cluster is driven by Brownian motion, however at the same time it is influenced by the overall magnetic dipole interaction of the cluster.
The energetically most favorable position for the particle to attach to the cluster is at either end of the chain.
Due to the high viscosity of the polymer, the diffusion coefficient of the particle is low.
As a consequence, the trajectory of the particle may be affected more efficiently by the dipolar interaction with the cluster, thus it is more likely moving towards the energetically most favorable position.
Although this argument applies to all cluster sizes, we observe a different power law for small clusters.
The fractal dimension $d = 1.74 \pm 0.01$ is close to the value for diffusion limited aggregation, DLA ($d = 1.715 \pm 0.004$).\cite{Witten1983,Pastor1995,Tolman1989}
Particle dynamics in diffusion limited aggregation are fully driven by Brownian motion.
Particles do not interact with each other unless they attach.
Once attached, they are permanently bound to the cluster.
For large cluster sizes, we see a dominating magnetic dipolar interaction between a nanoparticle and already existing clusters.
The magnetic dipolar interaction with a small cluster is weak.
Therefore, even the reduced diffusion coefficient is not enough to allow for the nanoparticle to be significantly influenced by the magnetic dipolar interaction.
2D Simulation also reproduce this effect.\cite{Pastor1995} 
Smaller cluster tend to exhibit a fractal dimension close to the DLA case.

Turning to the smaller nanoparticle FeNP5.
In measuring the magnetisation curves, we find that FeNP5 has no remanent magnetisation (see ESI), thus it is superparamagnetic.
As there is no magnetic driving force, they do not form chain-like aggregates. 
However, the fractal dimension of FeNP5 clusters is almost two, thus significantly higher as expected for DLA.
An explanation can be given by a restructuring process during aggregate formation. 
After attaching to the cluster, particles are able to slightly move towards the energetically most favorable position. 
For small clusters this leads to a significant compactification.\cite{Vicsek1984}
Here, the anisotropy of the particles may be crucial. 

The observation on aggregation can also explain the different compression behaviour of the two polymer nanocomposite films. Up to a surface pressure of \SI{15}{\milli\newton\per\metre} the isotherms for both composites are almost identical, thus the compression behaviour seems to be polymer dominated. The LB films investigated by AFM are prepared precisely at this point. For NPC10 the randomly oriented nanoparticle chains can be deformed thus leading to a weaker increase in surface pressure. 
The compact aggregates in NPC5 can not be further deformed. 
Their interaction with the polymer matrix is determined by the interaction of the oleic acid molecules and for the investigated concentration of particles compression behaviour is still dominated by the polymer.

\begin{figure}[t]
	\centering
	\includegraphics[width=0.6\textwidth]{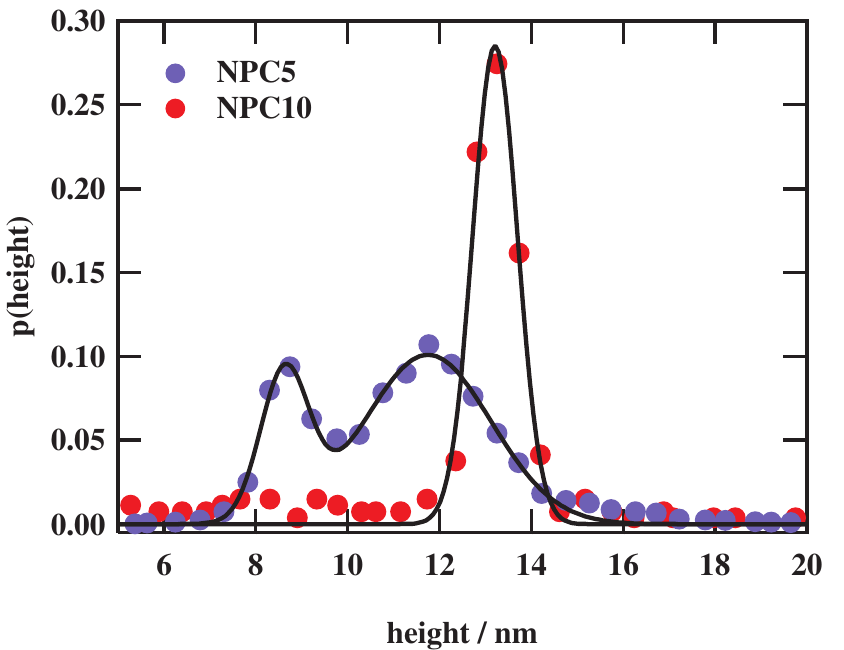}
	\caption{Height distribution of identified objects for NPC5 and NPC10 LB films. For the anisotropic FeNP5 a bimodal distribution is found, while FeNP10 exhibit a single maximum. Full lines represent guides for the eye.}
	\label{fig:AFM_npc_distribution}
\end{figure}

Another interesting aspect is the height distribution of the aggregates in the film.
We calculated the height distribution by extracting the maximum height for each object in both composites.
The same three images per particle size as for the fractal analysis were used.
Results for the normalized height distributions are shown in figure \ref{fig:AFM_npc_distribution}.
For NPC10, a single peak is observed with its maximum at \SI{13.3}{\nano\metre}.
This value is in very good agreement with the film thickness measured for the the pure FeNP10 films (see figure \ref{fig:fenp10}) and the particle size obtained by SAXS (see figure \ref{fig:fenp_saxs_formfactor}).
For NPC5, a bimodal distribution is observed with maxima at \SI{8.6}{\nano\metre} and \SI{11.7}{\nano\metre}.
Since there is no indication for buckling of the nanoparticle domains in the composite, the two maxima can be related to the spheroidal shape of the FeNP5 and represent flat and flipped configurations. The absence of buckling is not surprising because of the low concentration of nanoparticles in the film and the moderate surface pressure at LB film preparation. In contrast to NPC10 the positions of the maxima are shifted towards larger sizes in comparison to SAXS results. The difference is around \SI{4}{\nano\metre} which is comparable to the thickness of the polymer film. The in-situ XRR results indicate that the nanoparticles are distributed within the polymer film, however different surface affinities towards the silicon wafer may lead to a rearrangement, resulting in nanoparticles being pushed out of the polymer film. Since FeNP5 are smaller than the size of the polymer chain, this scenario seems more likely for these particles than for the larger ones.

\section{Conclusion}

In this study we investigated the structural properties of iron oxide nanoparticles at the air-water interface. Three different sizes of particles were characterised by small angle X-ray scattering (SAXS). All samples showed a narrow size distribution but only the two larger ones ($d_\text{FeNP10}=\SI{12.8}{\nano\metre}$, $d_\text{FeNP20}=\SI{21.4}{\nano\metre}$) are of spherical shape, while the smallest one is an oblate spheroid ($d_{\text{FeNP5}_\text{min}}=\SI{3.7}{\nano\metre}$, $d_{\text{FeNP5}_\text{maj}}=\SI{9.4}{\nano\metre}$). While FeNP20 particles do not form stable monolayers at the interface, the film structure and compression behaviour of FeNP5 and FeNP10 can be nicely characterised. At high surface pressures ($\Pi\leq\SI{18}{\milli\newton\per\metre}$) FeNP5 particles form a double layer, which may be facilitated by the anisotropic shape of these particles.

FeNP5 and FeNP10 can be introduced in a thin polymer matrix and form stable polymer nanocomposites at the air-water interface. 
XRR shows that nanoparticles are immersed in a \SI{3}{\nano\meter} to \SI{4}{\nano\meter} thick polymer layer at the air-water interface.
On LB films, the height distribution of aggregates seen by AFM reproduces the particle sizes measured by SAXS and XRR.
FeNP5 is superparamagnetic and forms small compact aggregates within transferred structures (LB film).
The formation of these aggregates can be explained by diffusion limited aggregation (DLA) in the limit of small cluster sizes with dominating restructuring effects.\cite{Vicsek1984} In contrast, FeNP10 is ferromagnetic and therefore forms open, chain-like aggregates due to an alignment of their magnetic moments along the chain. The existence of such randomly oriented chains of magnetic particles has been predicted by de Gennes theoretically even in the absence of an external magnetic field.\cite{deGennes1970}
Their existence has also been observed in magnetotactic bacteria as linear chains of magnetic nanoparticles coated with a lipid biomembrane.\cite{Ghaisari2017,Scheffel2005}
2D simulations find a fractal dimension of 1.13 for sufficiently large aggregates for particles with dominating magnetic interaction.\cite{Pastor1995,Holm2005} 
Although thermal energy should be dominating in this system, our results nicely agree with these predictions for large cluster sizes.
We explain this by the presence of the polymer matrix providing a highly viscous environment for the nanoparticles.
Therefore, nanoparticles are more easily guided towards the energetically most favorable position in the cluster.
For small cluster sizes the reduced diffusion coefficient of the particles can not compensate the small magnetic field provided by the aggregate.
Their fractal dimension is $d=1.74$, close to the DLA value.

\section{Acknowledgements}
The authors gratefully acknowledge Professor Dr. Oliver Gutfleisch and his group for performing the magnetisation measurements.

\bibliography{MyLibrary} 

\end{document}


\newpage
\section{Form Factors for SAXS Measurements}
For FeNP10 and FeNP20, we used a sphere form factor with a radius $r_c$ and scattering length density difference between particle and solvent $\Delta \rho_{cs}$:
\begin{equation}
	\label{eq:FormFactor}
	F_{\text{sphere}}(q,r_c)=r_c^3\Delta\rho_{cs}\frac{\sin(qr_c)-qr_c\cos(qr_c)}{(qr_c)^3}
\end{equation}
Polydispersity (PDI) of the radius is taken into account by a Schulz-Zimm distribution:
\begin{equation}
f(r,r_c)=\frac{1}{PDI^2}^\frac{1}{PDI^2}\left(\frac{r}{r_c}\right)^{\frac{1}{PDI^2}-1}\frac{\text{e}^{-\frac{r}{PDI^2r_c}}}{r_c\Gamma\left(\frac{1}{PDI^2}\right)}
     \label{eq:Schulz}
\end{equation}
Scattered intensity can be written as:
\begin{equation}
I(q) \propto \int\limits_0^{\infty}|F_{\text{sphere}}(q,r)|^2\cdot f(r,r_c)\,\text{d}r
\end{equation}

In case of FeNP5, a model for an oblate spheroid was used with minor radius $r_\text{minor}$, major radius $r_\text{major}$, scattering length density difference $\Delta\rho_{cs}$ and the first order Bessel function $J_1(x)$:
\begin{equation}
	\label{eq:FormFactor}
     F(q)=4 \pi \Delta\rho_{cs}r_\text{minor}r_\text{major}^2 
\ \frac{J_1(q u)}{qu}
\end{equation}
\begin{equation}
	u=\sqrt{r_\text{major}^2(1-\alpha^2)+r_\text{minor}^2\alpha^2}
\end{equation}
The orientation $\alpha$ of the oblate spheroid in respect to the scattering axis is limited by $0 \leq \alpha \leq 1$. Scattered intensity is given by an average over all possible orientation:
\begin{equation}
I(q)\propto\int\limits_0^1\left|F(q,u,\alpha)\right|^2\text{d}\alpha
\end{equation}
As a consequence, it is not possible to determine polydispersity of the two radii.
\newpage
\section{XRR and Langmuir Isotherms of FeNP20 Films}
\begin{figure}[!t]
	\centering
	\begin{minipage}{0.47\textwidth}
		\includegraphics[width=\textwidth]{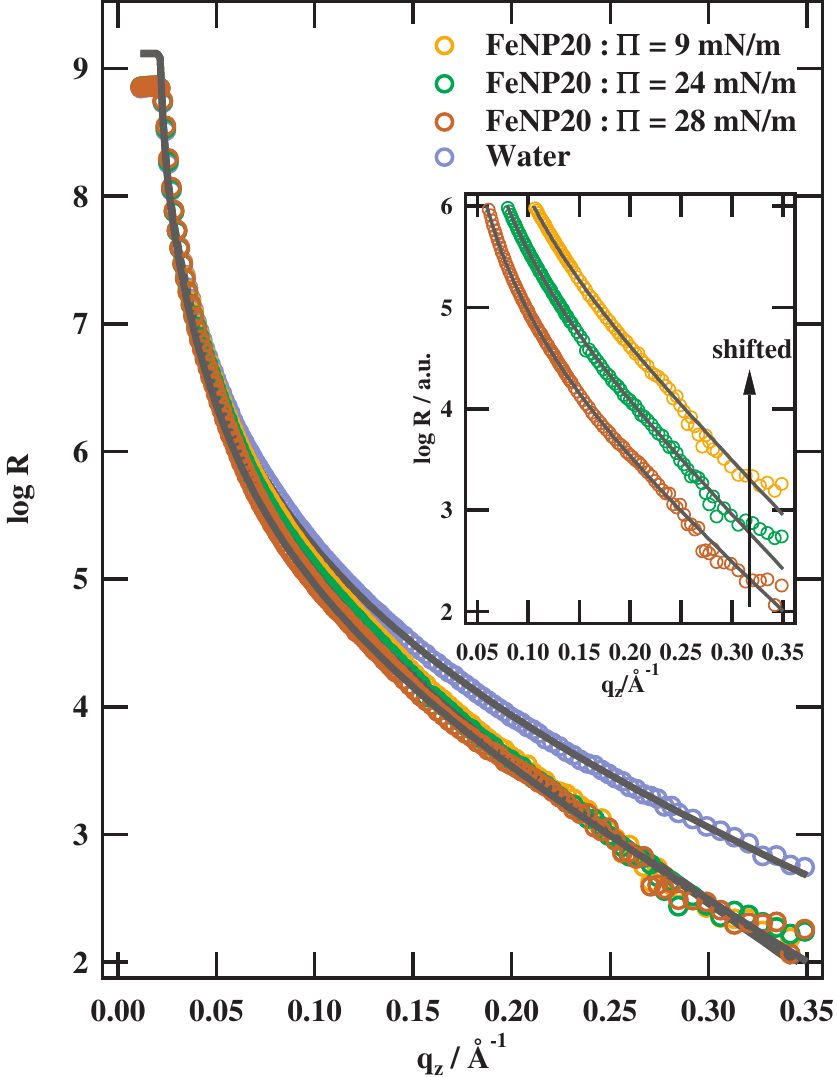}
	\end{minipage}
	\hspace{5mm}
	\begin{minipage}{0.47\textwidth}
		\includegraphics[width=0.96\textwidth]{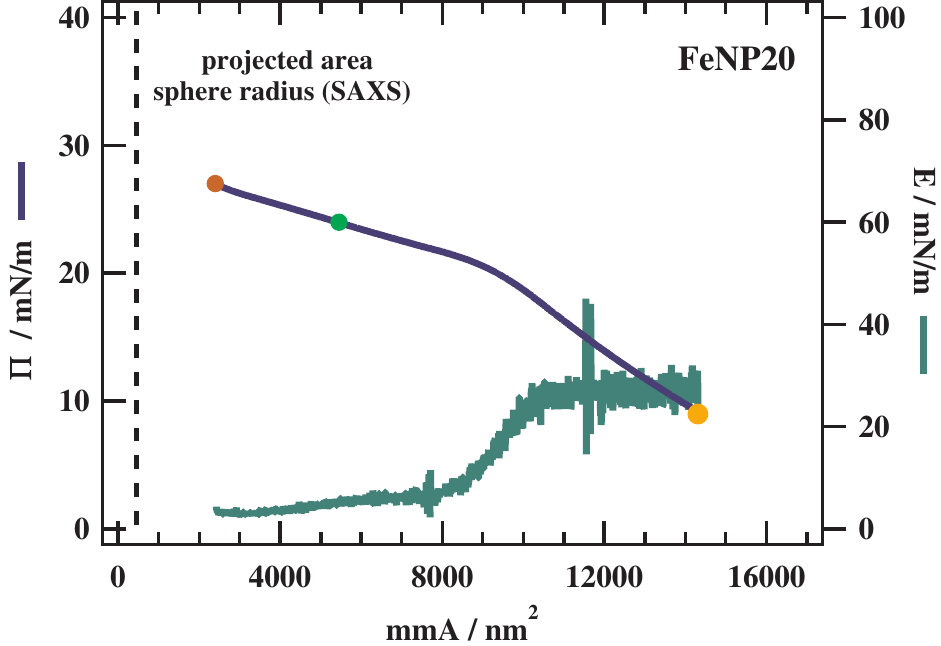}
		\includegraphics[width=0.85\textwidth]{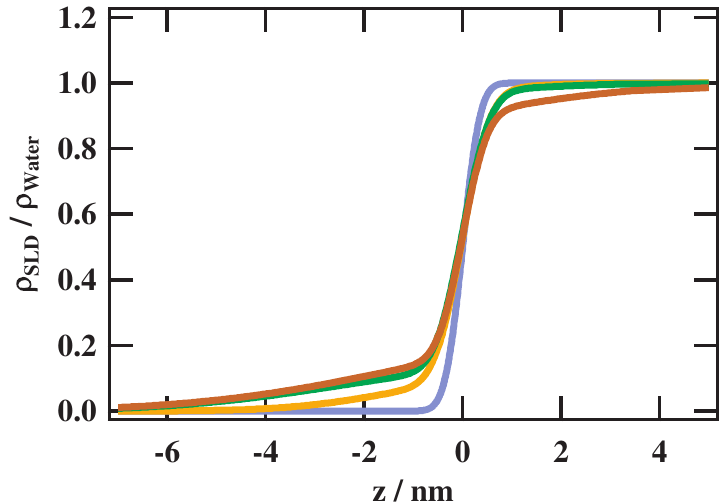}
	\end{minipage}
	\caption{X-Ray reflectivity curves (left hand side) measured for three different positions in the Langmuir isotherm (top panel right hand side) and the bare water surface prior to preparation of the FeNP20 film are shown. The Elastic modulus $E$ computed from the isotherm is depicted too. The bottom panel on the right hand side shows the normalised scattering length density profiles obtained from a one-layer model fitted to the raw data represented by the full lines in the reflectivity curves.}
	\label{fig:FeNP20}
\end{figure}

FeNP20 films were prepared at the air-water interface. The surface pressure is already significantly increased ($\Pi \approx 10\,\mathrm{mN/m}$) for mean molecular areas thirty times larger than the projected particle area measured by SAXS (see Langmuir isotherm in the top panel on the right hand side of figure \ref{fig:FeNP20}).
Reflectivity curves were measured for three different positions in the isotherm as highlighted in the isotherm.
The raw data shows essentially a rough surface (see left hand side of figure \ref{fig:FeNP20}). It can be fitted by a single layer of $2\,\mathrm{nm}$ with a roughness around $3\,\mathrm{nm}$.
The corresponding normalised scattering length density profiles are shown in the bottom panel on the right hand side of figure \ref{fig:FeNP20}.
There is no evidence for the formation of a single particle film for FeNP20. 
\newpage

\section{Polymer Nanocomposite Solutions in SAXS}
\begin{figure}[!t]
	\centering
	\includegraphics[width = 0.49\textwidth]{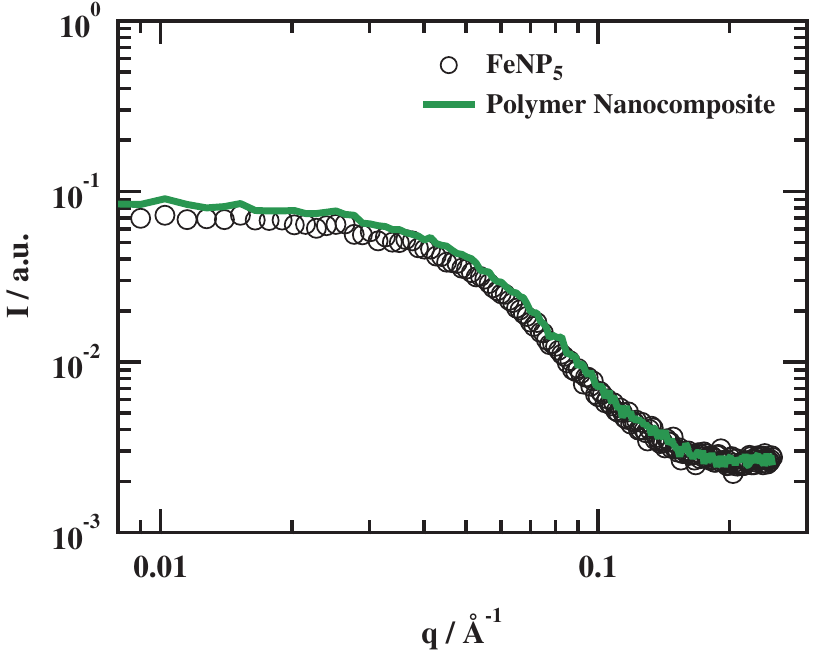}
	\hfill
	\includegraphics[width = 0.49\textwidth]{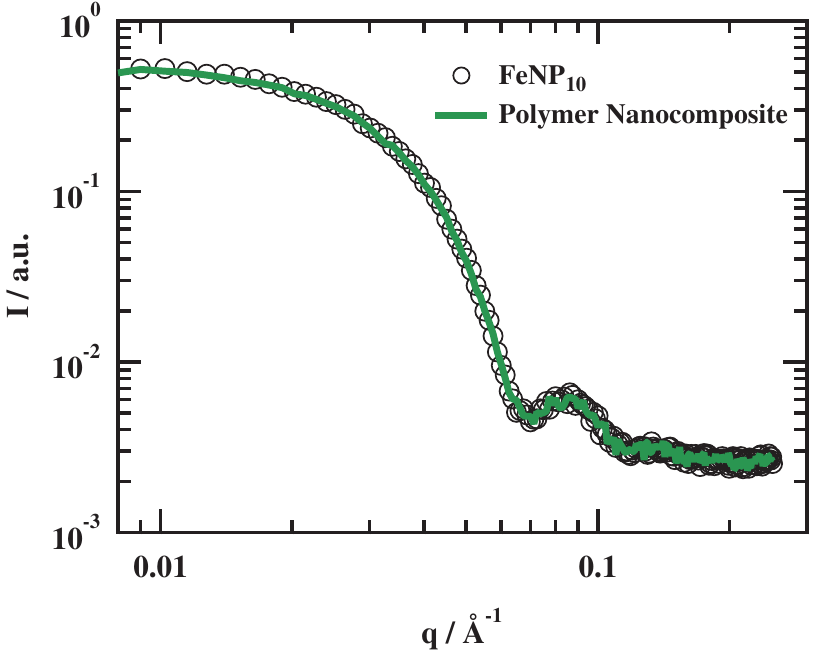}
	\caption{Small angle scattering data from polymer nanocomposites NPC5 (left hand side) and NPC10 (right hand side) dissolved in toluene. Pure particle solutions as shown in the main article are depicted for comparison. There is almost no difference between the curves, thus no evidence of aggregation can be found.}
	\label{NPC_FeNP_SAXS} 
\end{figure}

Polymer nanocomposite solutions were prepared as described in the experimental section of the main article. Mixtures of nanoparticles and polymer dissolved in toluene were measured by SAXS. 
As can be clearly seen in figure \ref{NPC_FeNP_SAXS}, obtained data is almost identical to pure nanoparticle solutions.
Thus, we conclude that prior to spreading no aggregation is present in the mixtures.

\newpage
\section{Off-specular Scattering for Polymer Nanocomposites}
\begin{figure}[!t]
	\centering
	\includegraphics[width=0.49\textwidth]{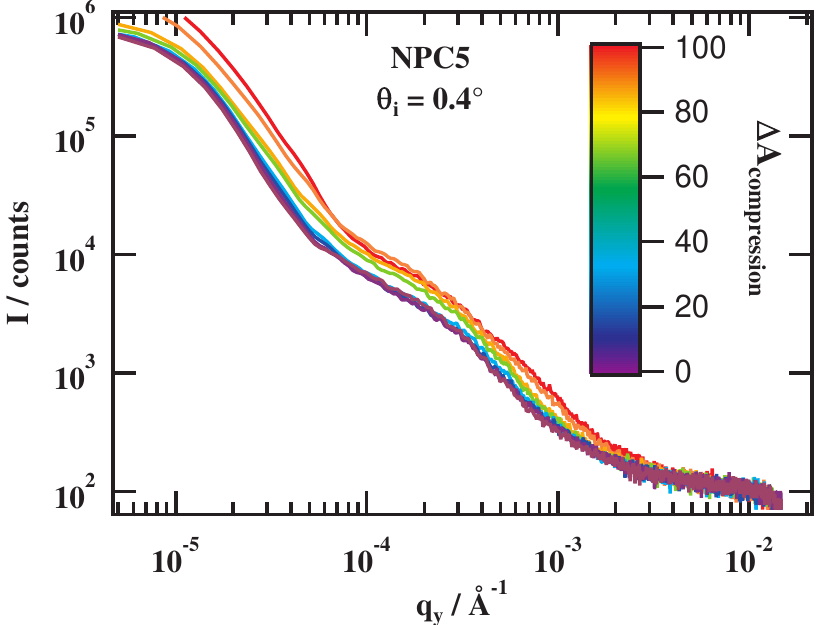}
	\hfill
	\includegraphics[width=0.49\textwidth]{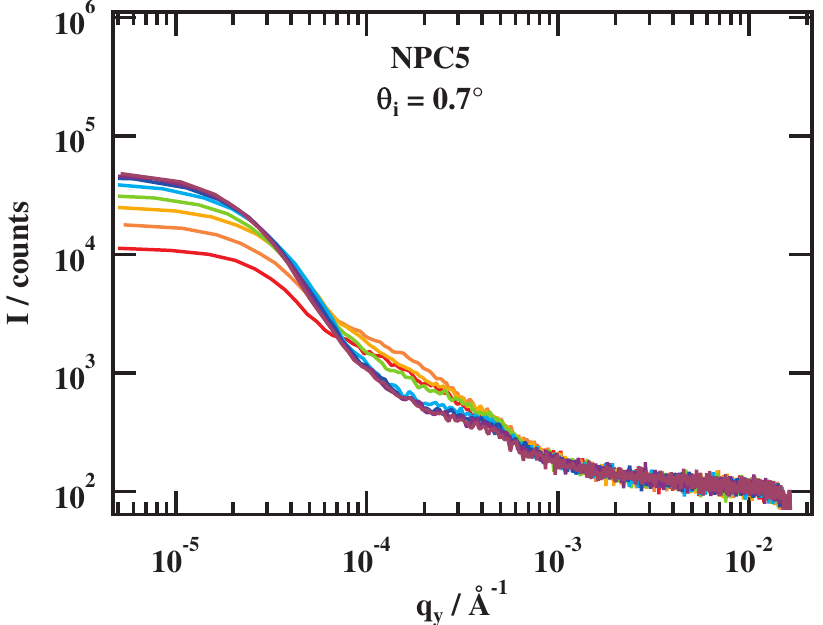}
	\includegraphics[width=0.49\textwidth]{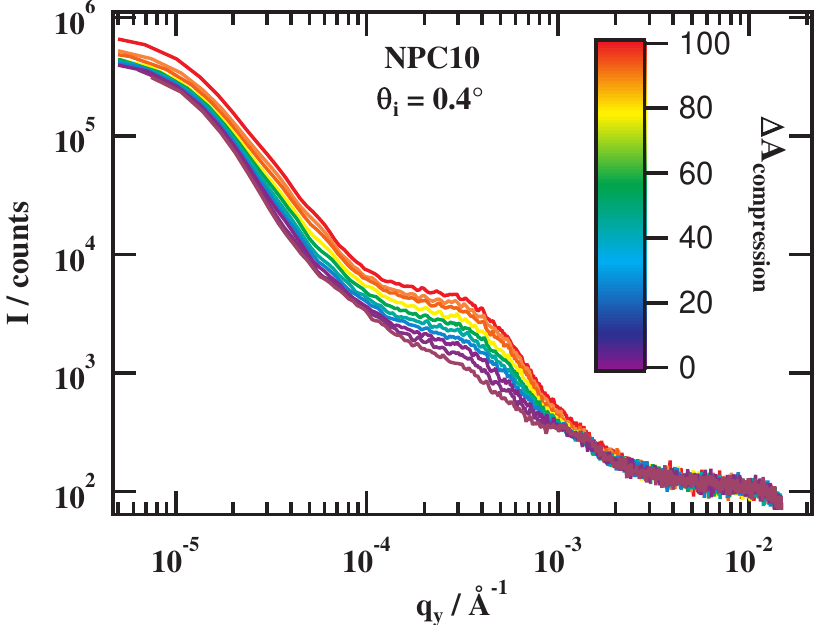}
	\hfill
	\includegraphics[width=0.49\textwidth]{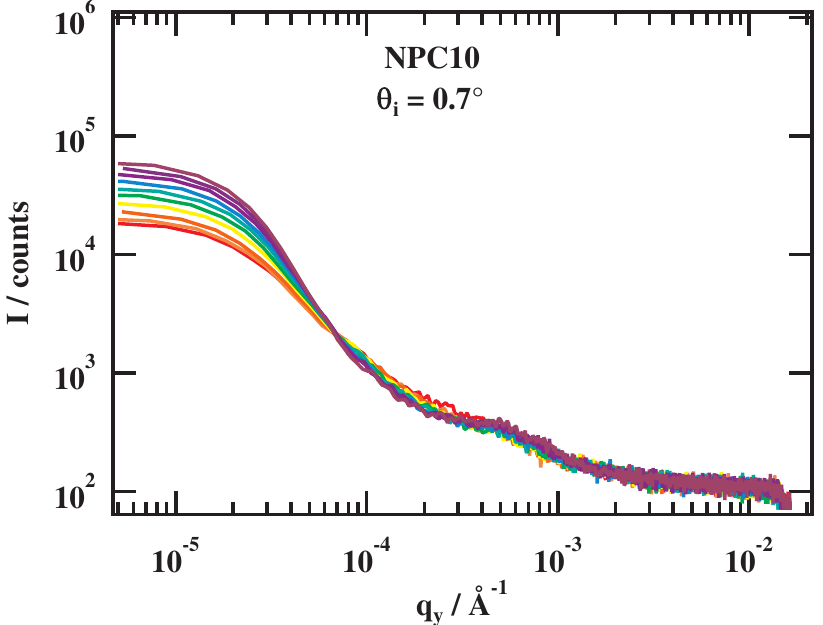}
	\caption{Off-specular intensity for NPC5 (top images) and NPC10 (bottom images) films vs. the in-plane scattering vector q$_{\text{y}}$. The incident angles are highlighted in the images. The colours refer to consecutive measuring points in the isotherm.}
	\label{NPC_offspecularscat} 
\end{figure}
Off-specular intensity was recorded for NPC5 and NPC10 films for subsequent measurement positions in the isotherm. Figure \ref{NPC_offspecularscat} shows the double logarithmic representation of intensity vs in-plane scattering vector q$_{\text{y}}=2\pi / \lambda \cdot(\cos(\theta_i)-\cos(\theta_f))$ for NPC5 \& NPC10 at two different incident angles $\theta_i$ of the X-ray tube. 
The colour scale represents the percentage of compression after the films were prepared at $\Pi \approx 18\,\mathrm{mN/m}$.
Scattering from lateral structures at the interface leads to a signal at a certain q$_{\text{y}}$, independent of the incident angle $\theta_i$.
On the other hand, scattering from bulk structures (such as the form factor of the particles) leads to a signal at a certain $\theta_i + \theta_f$, thus is not fixed to a certain q$_{\text{y}}$ when the incident angle is changed.
Both contributions are measured at the same time and can not be clearly separated.
This is shown in the comparison of left and right panel in figure \ref{NPC_offspecularscat}. 
Thus, it is impossible to investigate the surface scattering from the film by in-situ off-specular intensities.
The small peak visible for $\theta_i = 0.7\,^{\circ}$ at q$_{\text{y}}\approx 5\cdot 10^{-3}$ (right panel in figure \ref{NPC_offspecularscat}) is a scattering artifact of the experiment. It is visible in all spectra for $\theta_i + \theta_f \approx 1.8\,^{\circ}$ moving to smaller q$_{\text{y}}$ with increasing $\theta_i$.
\section{Magnetisation curves}
\begin{figure}[!t]
	\centering
	\includegraphics[width = 0.49\textwidth]{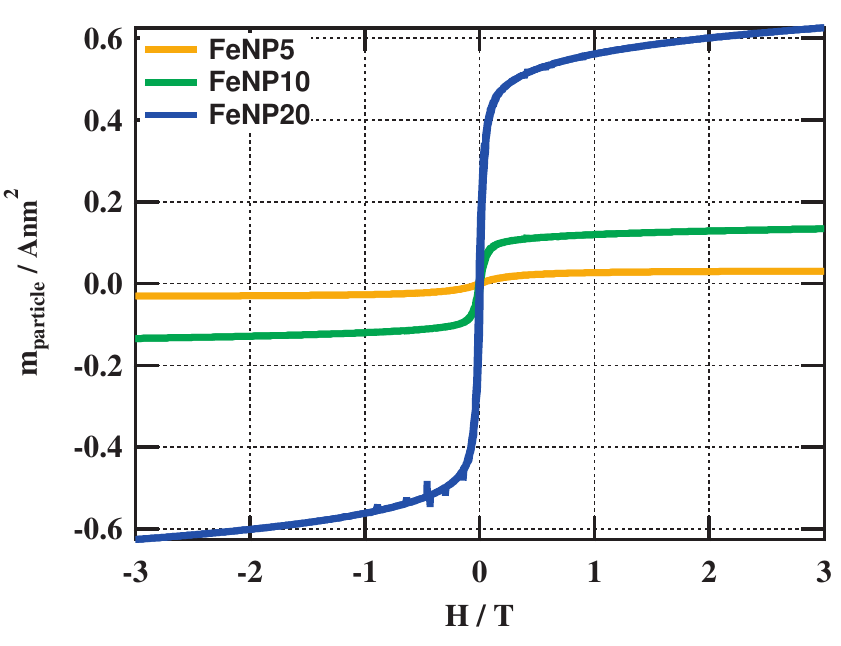}
	\hfill
	\includegraphics[width = 0.49\textwidth]{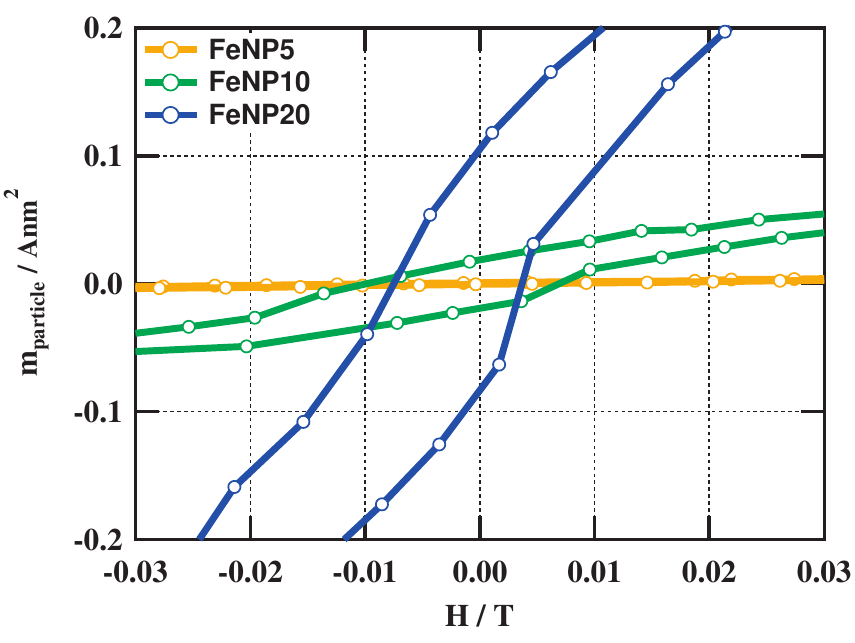}
	\caption{Magnetic hysteresis loops for all particle sizes with full curves shown in the left panel and zoomed around the center in the right panel.}
	\label{FENP_Magnetisation} 
\end{figure}
Isothermal magnetisation measurements were performed at room temperature using PPMS (Physical Properties Measurement System - Quantum Design PPMS-14) with VSM option (Vibrating Sample Magnetometer), under applied magnetic field up to \SI{3}{\tesla}. 
For this, the nanoparticle solutions were dried in a powder sample holder and fixed into a brass half tube holder (Quantum Design).

The left panel of figure \ref{FENP_Magnetisation} shows full magnetic hysteresis loops for all particle sizes. 
Zoomed curves around the center are shown in the right panel of the figure. It can clearly be seen that the remanent magnetisation vanishes with decreasing particle size. We obtain $M_\text{R} = \SI{0.094}{\ampere\nano\meter\squared}$ for FeNP20 and $M_\text{R} = \SI{0.019}{\ampere\nano\meter\squared}$ for FeNP10.

